\documentclass[envcountsame]{llncs}

\usepackage{multicol}
\usepackage{subfigure}
\usepackage{pstricks,pst-grad}
\usepackage{amssymb}
\usepackage{stmaryrd}
\usepackage{fix-cm}
\usepackage{amsmath}
\usepackage{color}
\usepackage{float}
\usepackage{wrapfig}
\usepackage{wasysym}
\usepackage{tikz}
\usetikzlibrary{arrows,shapes,snakes}
\usepackage{multirow}
\usepackage{subfigure}
\usepackage{float}
\usepackage{listings}
\usepackage{verbatim}
\usepackage[ruled,vlined]{algorithm2e}
\usepackage{pgf}
\usepackage{url}
\usepackage{tikz}
\usetikzlibrary{shapes,snakes}
\usetikzlibrary{positioning}
\usepackage[utf8]{inputenc}
\usepackage{etex}
\usepackage{bussproofs}
\usetikzlibrary{arrows,automata}
\usetikzlibrary{positioning}
\usetikzlibrary{arrows,shapes,fit,backgrounds}

\usepackage[normalem]{ulem}

\newcommand{\node}{\text{node}}

\newcommand{\abs}[1]{| #1 |}
\newcommand{\cyl}[1]{\text{Cyl}(#1)}

\newcommand{\tup}[1]{\langle #1 \rangle}

\definecolor{orange3}{rgb}{1.0,0.2538,0.1681}
\definecolor{blau}{rgb}{0,.39608,0.74118}
\definecolor{rot}{rgb}{0.79216,.12941,0.24706}
\definecolor{dgruen}{rgb}{0,.7,0}
\newcommand\Color[2]{\textcolor{#1}{#2}}
\newcommand{\nb}[1]{\marginpar{\scriptsize #1}}

\newcommand\Andreas[1]{\Color{rot}{(*#1*)}}
\newcommand\Javier[1]{\Color{dgruen}{(*#1*)}}
\newcommand\Stefan[1]{\Color{orange3}{(*#1*)}}
\newcommand\javier[1]{\nb{\Color{dgruen}{(*#1*)}}}
\newcommand\andreas[1]{\nb{\Color{rot}{(*#1*)}}}
\newcommand\stefan[1]{\nb{\Color{orange3}{(*#1*)}}}

 \renewcommand\Andreas[1]{}
 \renewcommand\Javier[1]{}
 \renewcommand\javier[1]{}
 \renewcommand\andreas[1]{}
 \renewcommand\stefan[1]{}
 \renewcommand\Stefan[1]{}

\usepackage{ifthen}
\newcommand{\techRep}{false} 
\newcommand{\iftechrep}{\ifthenelse{\equal{\techRep}{true}}}

\newcommand*{\Runs}{\text{Runs}}

\newcommand*{\Var}{\text{Var}}
\newcommand*{\Paths}{\text{Paths}}

\newenvironment{qtheorem}[2][]{%
{${}$\\[1mm]
\noindent \bf Theorem #2#1.}
\begin{itshape}%
}{%
\end{itshape}%
}

\newenvironment{qproposition}[2][]{%
{${}$\\[1mm]
\noindent\bf Proposition #2#1.}
\begin{itshape}%
}{%
\end{itshape}%
}

\newbox\subfigbox 
\makeatletter
{\def\caption##1{\gdef\subcapsave{\relax##1}}%
\let\subcapsave=\@empty 
\let\sf@oldlabel=\label
\def\label##1{\xdef\sublabsave{\noexpand\label{##1}}}%
\let\sublabsave\relax 
\setbox\subfigbox\hbox
\bgroup}
{\egroup 
\let\label=\sf@oldlabel
\subfigure[\subcapsave]{\box\subfigbox}}%
\makeatother
\newcommand{\nondet}{\text{nondet}}
\newcommand{\coin}{\text{coin}}
\newcommand{\lab}{\text{label}}

\newcommand{\Val}{\mathit{Val}}
\newcommand{\Resp}{\mathit{Resp}}

\renewcommand{\techRep}{false} 
\renewcommand{\iftechrep}{\ifthenelse{\equal{\techRep}{false}}}

\begin{document}
%
%
\pagestyle{headings}  

\author{Javier Esparza\inst{1} \and Andreas Gaiser\inst{1}\thanks{Andreas Gaiser is supported by the DFG Graduiertenkolleg 1480 (PUMA).} \and Stefan Kiefer\inst{2}\thanks{%
            Stefan Kiefer is supported by a postdoctoral fellowship
            of the German Academic Exchange Service (DAAD).}}
\title{
Proving Termination of Probabilistic Programs Using Patterns}
\institute{Institut f\"{u}r Informatik, Technische Universit\"{a}t M\"{u}nchen, Germany\\
    \texttt{\{esparza,gaiser\}@model.in.tum.de} \and
     Department of Computer Science, University of Oxford, United Kingdom\\
  \texttt{stefan.kiefer@cs.ox.ac.uk}}
\maketitle


\pagestyle{headings}  
\LinesNotNumbered

\begin{abstract}
Proving programs terminating is a fundamental computer science challenge.
Recent research has produced powerful tools that can check a wide range of programs for termination.
The analog for probabilistic programs, namely termination with probability one (``almost-sure termination''),
 is an equally important property for randomized algorithms and probabilistic protocols.
We suggest a novel algorithm for proving almost-sure termination of probabilistic programs.
Our algorithm exploits the power of state-of-the-art model checkers and termination provers for nonprobabilistic programs:
 it calls such tools within a refinement loop
 and thereby iteratively constructs a ``terminating pattern'',
  which is a set of terminating runs with probability~one.
We report on various case studies illustrating the effectiveness of our algorithm.
As a further application, our algorithm can improve lower bounds on reachability probabilities.
\end{abstract}

\section{Introduction}

\newcommand{\ltl}[1]{\mathbf{#1}\,}

Proving program termination is a fundamental challenge of computer science.
Termination is expressible in temporal logic, and so checkable in principle by LTL or
CTL model-checkers. However, recent research has shown that special purpose tools, like
Terminator and ARMC~\cite{armc,cook}, and techniques like \emph{transition invariants}, can be dramatically
more efficient \cite{DBLP:conf/lics/PodelskiR04,phd_andrey,DBLP:conf/tacas/PodelskiR11}.

The analog of termination for probabilistic programs is termination with probability one, or \emph{almost sure termination},
 abbreviated here to \emph{a.s.-termination}.
Since a.s.-termination is as important for randomized algorithms and probabilistic protocols as termination is for regular programs, the
question arises whether the very strong advances in automatic termination proving termination can be exploited in the probabilistic case.
However, it is not difficult to see that, without further restricting the question, the answer is negative.
The reason is that termination is a purely topological property of the transition system associated to the program, namely absence of cycles,
 but a.s.-termination is not. Consider for instance the program

\begin{center}
\begin{minipage}{10cm}
\begin{verbatim}
k = 1;  while (0 < k) { if coin(p) k++ else k--}
\end{verbatim}
\end{minipage}
\end{center}

\noindent where \verb+coin(p)+ yields $1$ with probability $0 < p < 1$, and $0$ with probability $(1-p)$.
The program has the same executions for all values of $p$ (only their probabilities change), but it only terminates a.s.\ for $p \leq 1/2$.
This shows that proving a.s.-termination requires arithmetic reasoning not offered by termination provers.

The situation changes if we restrict our attention to \emph{weakly finite}
probabilistic programs. Loosely speaking, a program is weakly finite if the set of states reachable from any initial state is finite.
Notice that the state space may be infinite,
 because the set of initial states may be infinite.
Weakly finite programs are a large class, which in particular contains
\emph{parameterized probabilistic programs}, i.e., programs with parameters that can be initialized to arbitrary large values,
 but are finite-state for every valuation of the parameters. \andreas{Bemerkung 3: ... ``We guess you mean that after fixing the value of the parameter
our programs always have a finite-state space. We'll clarify this.'' Das was hier steht, muesste eigtl reichen, oder?}
One can show that a.s.-termination is a topological property for weakly finite programs. If the
program is deterministic, then it terminates a.s.\ if{}f for every reachable state~$s$ there is a path
\textcolor{black}{in the non-probabilistic program obtained by making all
probabilistic choices nondeterministic}
leading from $s$ to a terminating state, which corresponds to the CTL property $AG \, EF \,{\it end}$.
\andreas{Halbsatz (in rot) hinzugefuegt, reicht das?}
\stefan{Ich glaub, das Problem ist, dass nicht klar ist, was deterministic bedeutet. Das wird hier zum ersten Mal verwendet,
 und koennte als nicht-probabilistisch missverstanden werden.}

(In the nondeterministic case there is also a corresponding topological property.) As in the nonprobabilistic case,
 generic infinite-state model checkers perform poorly
for these properties because of the quantifier alternation
$AG \, EF$. In particular, CEGAR approaches usually fail, because, crudely speaking,
they tend to unroll loops, which is essentially useless for proving termination.

In \cite{DBLP:conf/fossacs/AronsPZ03}, Arons, Pnueli and Zuck present a different and very
elegant approach that reduces \emph{a.s.-termination}
of a \emph{probabilistic} program to \emph{termination} of a \emph{nondeterministic} program
obtained with the help of a {\it Planner}. A Planner occasionally and infinitely often determines the outcome of the next $k$ random choices for some fixed $k$,
 while the other random choices are performed nondeterministically.
 In this paper we revisit and generalize this approach, with the goal
of profiting from recent advances on termination tools and techniques not available when
\cite{DBLP:conf/fossacs/AronsPZ03} was published. While we also partially fix the outcome of
random choices, we do so more flexibly with the help of \emph{patterns}. A first advantage of
patterns is that we are able to obtain a \emph{completeness} result for weakly finite programs,
which is not the case for Planners. Further, in contrast to \cite{DBLP:conf/fossacs/AronsPZ03},
we show how to automatically derive patterns
for finite-state and weakly finite programs using an adapted version of the CEGAR approach.
Finally, we apply our a.s.-termination technique to improve CEGAR-algorithms
for \emph{quantitative} probabilistic verification \cite{wachter,hermanns,kattenbelt,DBLP:conf/sas/EsparzaG11}.

In the rest of this introduction we explain our approach by means of examples. First we discuss
finite-state programs and then the weakly finite case.

\paragraph{Finite-state programs.} Consider the finite-state program \texttt{FW} shown on the left of Fig.~\ref{fig:firewire}.
It is an abstraction of part of the FireWire protocol~\cite{mciver}.
\begin{figure}[ht]
\begin{minipage}{3,5cm}
\begin{verbatim}
k = 0;
while (k < 100) {
  old_x = x;
  x = coin(p);
  if (x != old_x) k++
}
\end{verbatim}
\end{minipage}
\qquad
\begin{minipage}{6cm}
\begin{verbatim}
c1 = ?; c2 = 2;
k = 0;
while (k < 100) {
  old_x = x;
  if (c1 > 0)      {x = nondet(); c1--}
  elseif (c2 = 2 ) { x = 0; c2--}
  elseif (c2 = 1 ) { x = 1; c2--}
  else /* c1 = 0 and c2 = 0 */ {c1 = ?; c2 = 2}}
  if (x != old_x) k++
}
\end{verbatim}
\end{minipage}
\caption{The programs \texttt{FW} and \texttt{FW'}.}
\label{fig:firewire}
\end{figure}
Loosely speaking, \texttt{FW} terminates a.s.\ because if we keep tossing a coin then
with probability 1 we observe 100 times two consecutive tosses with the opposite outcome (we even see
100 times the outcome $01$). More formally, let $C = \{0,1\}$, and let
us identify a {\em run} of \texttt{FW} (i.e., a terminating or infinite execution) with the sequence of $0$'s and $1$'s corresponding to the results of the coin tosses carried out during it.
For instance, $(01)^{51}$ and $(001100)^{50}$ are terminating runs of \texttt{FW}, and $0^\omega$ is a nonterminating run.
\texttt{FW} terminates because the runs that are prefixes of $(C^* 01)^\omega$ have probability 1, and all
of them terminate. But it is easy to see that these are also the runs of the nondeterministic program \texttt{FW'} on the right of Fig.~\ref{fig:firewire}
where \verb+c = ?+ nondeterministically sets \verb+c+ to an arbitrary nonnegative integer.
Since termination of \texttt{FW'} can easily be proved with the help
of ARMC, we have proved a.s.-termination of \texttt{FW}.

\andreas{Javiers Proof Rule: Wenn sie drinbleibt, dann wo, und in welcher Form? Problem: Begriff run (Javier identifiziert ihn mit
seiner
coin toss sequence zuvor)}
\textcolor{black}{
Our reasoning is based on the following simple proof rule, with $P$ a probabilistic program
and $R$ a set of runs of $P$:
\begin{prooftree}
\AxiomC{$Pr[R] = 1$}
\AxiomC{Every $r \in R$ is terminating}
\BinaryInfC{$P$ terminates a.s.}
\end{prooftree}
}
\stefan{Ich denke, das sollte am besten schon in den Abschnitt mit dem Planner eingebaut werden.}

We present an automatic procedure leading from \texttt{FW} to \texttt{FW'} based on the notion
of {\em patterns}. A pattern is a subset of $C^\omega$ of the form $C^* w_1 C^* w_2 C^* w_3 \ldots$, where $w_1, w_2 , \ldots \in C^*$.
We call a pattern \emph{simple} if it is of the form $(C^* w)^\omega$.
A pattern~$\Phi$ is \emph{terminating (for a probabilistic program $P$)} if all runs of $P$ that
\emph{conform} to~$\Phi$, i.e., that are prefixes of words of $\Phi$, terminate. In the paper we
 prove the following theorems:
\begin{itemize}
\item[(1)] For every pattern $\Phi$ and program $P$, the $\Phi$-conforming runs of $P$ have
probability 1.
\item[(2)] Every finite-state program has a simple terminating pattern.
\end{itemize}
\noindent By these results, we can show that \texttt{FW} terminates a.s.\
by finding a simple terminating pattern $\Phi$, taking for $P'$
a nondeterministic program whose runs are the $\Phi$-conforming runs of $P$, and proving that
$P'$ terminates. In the paper we show how to automatically find $\Phi$ with the help of
a finite-state model-checker (in our experiments we use SPIN).
We sketch the procedure using \texttt{FW} as example. First we check if
some run of \texttt{FW} conforms to $\Phi_0 = C^\omega$, i.e., if some run of \texttt{FW}
 is infinite, and get $v_1= 0^\omega$ as answer. Using an algorithm provided in the paper, we compute
 a {\em spoiler} $w_1$ of $v_1$: a finite word that is not an infix of $v_1$.
The algorithm yields $w_1 = 1$. We now check if some run of \texttt{FW} conforms to
$\Phi_1 = (C^* w_1)^\omega$, and get $v_2 = 1 ^\omega$ as counterexample, and construct
 a spoiler $w_2$ of \emph{both} $v_1$ and~$v_2$: a finite word that is an infix of \emph{neither}
$v_1^\omega$ \emph{nor} $v_2^\omega$. We get $w_2 = 01$, and check if some run of \texttt{FW}
conforms to $\Phi_2 = (C^* w_2)^\omega$. The checker finds no counterexamples,
and so $\Phi_2$ is terminating. In the paper we prove that the procedure is complete, i.e., produces
a terminating pattern for any finite-state program that terminates a.s.

\paragraph{Weakly finite programs.} We now address the main goal of the paper:
proving a.s.-termination for weakly finite programs. Unfortunately, Proposition (2) no longer holds.
Consider the random-walk program \texttt{RW} on the left of Fig.~\ref{fig:rw}, where $N$ is an input variable.
\begin{figure}
\begin{minipage}{4cm}
\begin{verbatim}
k = 1;
while (0 < k < N) {
  if coin(p) k++ else k--
}
\end{verbatim}
\end{minipage}
\qquad
\begin{minipage}{6cm}
\begin{verbatim}
K = 2; c1 = ?; c2 = K;
k = 1
while (0 < k < N) {
    if (c1 > 0) {
      if nondet() k++ else k--; c1--
    };
    elseif (c2 > 0) {k--; c2--}
    else {K++; c1 = ?; c2 = K}
}
\end{verbatim}
\end{minipage}
\caption{The programs \texttt{RW} and \texttt{RW'}}
\label{fig:rw}
\end{figure}
\texttt{RW} terminates a.s., but we can easily show (by setting $N$ to a large enough value) that no simple pattern is terminating.
However, there \emph{is} a terminating pattern, namely
$\Phi=C^* 00 C^* 000 C^* 0000 \ldots$: every $\Phi$-conforming run terminates, whatever value $N$ is set to. Since, by result (1), the $\Phi$-conforming runs
have probability 1 (intuitively, when tossing a coin we will eventually see longer and longer
chains of $0$'s), \texttt{RW} terminates a.s.
In the paper we show that this is not a coincidence by proving the following completeness result:
\begin{itemize}
\item[(3)] Every weakly finite program has a (not necessarily simple) terminating pattern.
\end{itemize}
\noindent In fact, we even prove the existence of a \emph{universal} terminating pattern, i.e., a single pattern
$\Phi_u$ such that for all weakly finite, a.s.-terminating probabilistic programs all $\Phi_u$-conforming runs
terminate. This gives a universal reduction of a.s.-termination to termination, but one that is not very useful in
practice. In particular, since the universal pattern \emph{is} universal, it is not tailored towards making
the proof of any particular program simple. For this reason we propose a technique that reuses the procedure
for finite-state programs, and extends it with an extrapolation step in order to produce a candidate for
a terminating pattern.  We sketch the procedure using \texttt{RW} as example. Let \texttt{RW}$_i$ be the
program \texttt{RW} with $N=i$. Since every \texttt{RW}$_i$ is finite-state,
 we can find terminating patterns $\Phi_i = (C^* u_i)^\omega$ for a finite set of values of $i$, say for $i=1,2,3,4,5$.
We obtain $u_1 = u_2 = \epsilon$, $u_3 = 00$, $u_4 = 000$, $u_5 = 000$. We prove in the paper that
$\Phi_i$ is not only terminating for \texttt{RW}$_i$, but also for every \texttt{RW}$_j$ with
$j \leq i$. This suggests to extrapolate and take the pattern
$\Phi = C^* 00 C^*000 C^* 0000 \ldots$ as a candidate for a terminating pattern for \texttt{RW}.
We automatically construct the nondeterministic program \texttt{RW'} on the right of Fig.~\ref{fig:rw}.
Again, ARMC proves that \texttt{RW'} terminates, and so that \texttt{RW} terminates a.s.

\smallskip\emph{Related work.}
A.s.-termination is highly desirable for protocols if termination within a fixed number of steps is not feasible.
For instance, \cite{Bracha:1985} considers the problem of reaching consensus within a set of interconnected processes,
some of which may be faulty or even malicious.
They succeed in designing a probabilistic protocol to reach consensus a.s.,
although it is known that no deterministic algorithm terminates within a bounded number of steps.
A well-known approach for proving a.s.-termination are Pnueli et al.'s notions of extreme fairness and
$\alpha$-fairness~\cite{DBLP:conf/stoc/Pnueli83,PnueliProbVerification}. These proof methods, although complete for finite-state systems,
are hard to automatize and require a lot of knowledge about the considered program. The same applies for the approach of
McIver et al. in~\cite{McIverPGCL} that offers proof rules for probabilistic loops in pGCL, an extension of Dijkstra's guarded language.
The paper~\cite{MonniauxSAS01} discusses probabilistic termination in an abstraction-interpretation framework.
It focuses on programs with a (single) loop and proposes a method of proving
that the probability of taking the loop $k$ times decreases exponentially with~$k$.
This implies a.s.-termination. In contrast to our work there is no tool support in~\cite{MonniauxSAS01}.

\smallskip\emph{Organization of the paper.}
Sections 2 contains preliminaries and the syntax and semantics of our model of probabilistic programs.
Section 3 proves soundness and completeness results for termination of weakly finite programs.
Section 4 describes the iterative algorithm for generating patterns.
Section 5 discusses case studies.
Section 6 concludes.
For space reasons, a full discussion of nondeterministic programs and some missing proofs
\iftechrep{have been moved to an appendix.}
{are omitted. They can be found in the full version of the paper in~\cite{patternTechRep}.}
\iftechrep{A shorter version of this paper will appear in the proceedings of the \emph{24th Computer Aided Verification conference (CAV 2012)}.}

\section{Preliminaries} \label{sec:preliminary}

For a finite nonempty set~$\Sigma$,
 we denote by $\Sigma^*$ and $\Sigma^\omega$ the sets of finite and
  infinite words over~$\Sigma$, and set $\Sigma^\infty = \Sigma^* \cup \Sigma^\omega$.


\subsubsection*{Markov Decision Processes and Markov chains.}
A {\em Markov Decision Process} (MDP) is a tuple $\mathcal{M} = (Q_A, Q_P, \text{Init}, \mathord\rightarrow, \text{Lab}_A, \text{Lab}_P)$,
 where $Q_A$ and $Q_P$ are countable or finite sets of \emph{action nodes} and \emph{probabilistic nodes},
 $\text{Init} \subseteq Q_A \cup Q_P$ is a set of \emph{initial nodes}, and
 $\text{Lab}_A$ and $\text{Lab}_P$ are disjoint, finite sets of {\em action labels} and {\em probabilistic labels}. Finally, the relation~$\mathord\rightarrow$ is equal to $\mathord{\rightarrow_A} \cup \mathord{\rightarrow_P}$,
  where $\mathord{\rightarrow_A} \subseteq Q_A \times \text{Lab}_A \times (Q_A \cup Q_P)$ is a set of {\em action transitions},
  and $\mathord{\rightarrow_P} \subseteq Q_P \times (0,1] \times \text{Lab}_P \times Q$ is a set of {\em probabilistic transitions}
   satisfying the following conditions: 
    (a) if $(q,p,l,q')$ and $(q, p',l,q')$ are probabilistic transitions, then $p = p'$;
    (b) the probabilities of the outgoing transitions of a probabilistic node add up to~$1$.
We also require that every node of~$Q_A$ has at least one successor in~$\mathord{\rightarrow_A}$.
If $Q_A = \emptyset$ and $\text{Init} = \{q_I\}$ then we call $\mathcal{M}$ a \emph{Markov chain} and write
 $\mathcal{M} = (Q_P, q_I, \mathord\rightarrow,\text{Lab}_P)$.

We set $Q = Q_A \cup Q_P$ and $\text{Lab} = \text{Lab}_A \cup \text{Lab}_P$.
We write $q \xrightarrow l q'$ for $(q, l, q') \in \mathord{\rightarrow_A}$,
and $q \xrightarrow {l,p} q'$
for $(q, p, l, q') \in \mathord\rightarrow_P$ (we skip $p$ if it is irrelevant).
For $w =l_1l_2\ldots l_n \in \text{Lab}^*$, we write $q \xrightarrow {w} q'$ if there exists a path
 $q = q_0 \xrightarrow {w_1} q_1 \xrightarrow {w_2} \ldots \xrightarrow {w_n} q_n = q'$.


\begin{figure}[tbp]
\centering
 \tikzset{
   dim/.style={
           rounded corners,
           minimum height=2em,
           inner sep=2pt,
           text centered,
           },
    state/.style={
           rectangle,
           rounded corners,
           draw=black, very thick,
           minimum height=2em,
           inner sep=2pt,
           text centered,
           },
   act/.style={
           circle,
           draw=black, very thick,
           minimum height=2em,
           inner sep=2pt,
           text centered,
           },
  fail/.style={circle, double, draw=black, inner sep=2pt, minimum height=2em, very thick},
  nothing/.style={
           rectangle,
           rounded corners,
           minimum height=2em,
           inner sep=0pt,
           text centered,
           }
  }
\scalebox{0.65}{
\begin{tikzpicture}[->,>=stealth',scale=1,font=\Large]

 \node[dim, anchor=center] (S1)
 {$ $};

 \node[act] (A1) [right =of S1]
 {$q_a$};

 \node[state] (P1) [right =1.5cm of A1]
 {$q_1$};

 \node[state] (P2) [right =2cm of P1]
 {$q_2$};

 \node[state] (P3) [right =1.5cm of P2]
 {$q_3$};

 \path[very thick]
       (S1) edge[above] node [above] {$ $} (A1)
       (A1) edge[above] node [below] {$a_1$} (P1)
       (P1) edge[above] node [below] {$\tup{\tau, 1}$} (P2)
       (P2) edge[bend right =65] node [above] {$\tup{c_0, \frac{1}{2}}$} (P1)
       (P2) edge[above] node [below] {$\tup{c_1, \frac{1}{2}}$} (P3)
       (P3) edge[loop right] node [loop right] {$\tup{\tau, 1}$} (P3)
       (A1) edge[bend left =65] node [above] {$a_0$} (P3);
\end{tikzpicture}
}
\caption{Example MDP.}\label{fig:example1}
\end{figure}
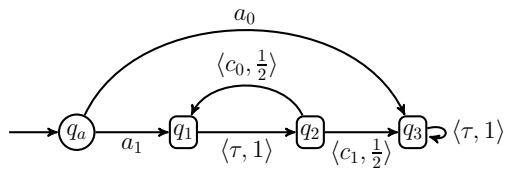

\begin{example}
Figure~\ref{fig:example1} shows an example of a Markov Decision Process
 $\mathcal{M} = (\{q_a\}, \{q_1,q_2,q_3\}, \text{Init}, \mathord\rightarrow, \text{Lab}_A, \text{Lab}_P)$,
  with action labels $a_0, a_1$, probabilistic labels $\tau, c_0, c_1$, and a single initial node $q_a$.
\end{example}

\subsubsection*{Runs, paths, probability measures, traces.}
A \emph{run} of an MDP $\mathcal{M}$ is an infinite word $r = q_0 l_0 q_1 l_1 \ldots\in (Q\text{Lab})^\omega$ such that
 for all $i \geq 0$ either $q_i \xrightarrow {l_i,p} q_{i+1}$ for some $p \in (0,1]$ or
$q_i \xrightarrow {l_i} q_{i+1}$. We call the run \emph{initial} if $q_0 \in \text{Init}$.
We denote the set of runs starting at a node $q$ by $\Runs^{\mathcal{M}}(q)$, and the set of all
runs starting at initial nodes by $\Runs(\mathcal{M})$.

A {\em path} is a proper prefix of a run. We denote by $\Paths^{\mathcal{M}}(q)$ the set of all paths starting at~$q$.
We often write $r = q_0 \xrightarrow {l_0} q_1 \xrightarrow {l_1} q_2 \xrightarrow {l_2} \ldots$ instead of $r = q_0 l_0 q_1 \ldots$ for both runs and paths,
and skip the superscripts of $\Runs(\cdot)$ and $\Paths(\cdot)$ if the context is clear.

We take the usual, cylinder-based definition of a probability measure~$\Pr_{q_0}$ on the set
of runs of a Markov chain~$\mathcal{M}$ starting at a state~$q_0 \in \text{Init}$
(see e.g.~\cite{principlesMC} or \iftechrep{the appendix}{\cite{patternTechRep}}) for details).
For general MDPs, we define a probability measure~$\Pr^S_{q_0}$ with respect to a
 \emph{strategy}~$S$. We may drop the subscript if the initial state is irrelevant or understood.


The {\em trace} of a run $r = q_0 \xrightarrow {\alpha_0} q_1 \xrightarrow {\alpha_1} \ldots \in \Runs(\mathcal{M})$, denoted by $\bar{r}$,
 is the infinite sequence $\alpha_0 \alpha_1 \ldots \in \text{Lab}$ of labels.
Given $\Sigma \subseteq \text{Lab}$, we define $\bar{r}|_\Sigma$
as the projection of $\bar{r}$ onto~$\Sigma$. Observe that $\bar{r}|_\Sigma$ can be finite.

\subsection{Probabilistic Programs}

We model probabilistic programs as flowgraphs whose transitions are labeled
with {\em commands}. Since our model is standard and very similar to~\cite{kattenbelt},
we give an informal but hopefully precise enough definition.
Let $\Var$ be a set of variable names over the integers (the variable domain could be easily extended),
 and let $\Val$ be the set of possible {\em valuations} of $\Var$, also called \emph{configurations}. The set of commands contains
\begin{itemize}
\item conditional statements, i.e., boolean combinations of expressions
$e \leq e'$, where $e,e'$ are arithmetic expressions (e.g, $x+y \leq 5 \wedge y \geq 3$);
\item deterministic assignments $x := e$ and nondeterministic assignments $x := \nondet()$ that
nondeterministically assign to $x$ the value $0$ or $1$;
\item  probabilistic assignments $x := \coin(p)$ that assign to $x$ the value $0$ or $1$ with probability $p$ or $(1-p)$, respectively.
\end{itemize}

A \emph{probabilistic program} $P$ is a tuple $(\mathcal{L}, I, \mathord{\hookrightarrow}, \lab, \bot, \top)$,
 where $\mathcal{L}$ is a finite set of control flow {\em locations},
 $I \subseteq \Val$ is a set of {\em initial configurations},
 $\mathord{\hookrightarrow} \subseteq \mathcal{L} \times \mathcal{L}$ is the {\em flow relation}
  (as usual we write $l \hookrightarrow l'$ for $(l,l') \in \mathord{\hookrightarrow}$, and call the elements of $\mathord{\hookrightarrow}$ {\em edges}),
 $\lab$ is a function that assigns a command to each edge, $\bot$ is the {\em start location}, and $\top$ is the {\em end location}.
The following standard conditions must hold:
 ($i$) the only outgoing edge of $\top$ is $\top \hookrightarrow \top$;
 ($ii$) either all or none of the outgoing edges of a location are labeled by conditional statements;
  if all, then every configuration satisfies the condition of exactly one outgoing edge;
  if none, then the location has exactly one outgoing edge;
 ($iii$) if an outgoing edge of a location is labeled by an assignment, then it is the only outgoing edge of this location.

A location is {\em nondeterministic} if it has an outgoing edge labeled by a nondeterministic assignment, otherwise it is {\em deterministic}.
Deterministic locations can be {\em probabilistic} or {\em nonprobabilistic}.
A program is deterministic if all its locations are deterministic.

\subsubsection{Program Semantics.}

The semantics of a probabilistic program is an MDP.
Let $P$ be a \emph{probabilistic program} $(\mathcal{L}, I, \mathord{\hookrightarrow}, \lab, \bot, \top)$,
 and let $\mathcal{L}_D, \mathcal{L}_A$ denote the sets of deterministic and nondeterministic locations of~$P$.
The semantics of $P$ is the MDP $\mathcal{M}_P := (Q_A, Q_D, \text{Init}, \rightarrow, \text{Lab}_A, \text{Lab}_P)$, where
$Q_A = \mathcal{L}_A \times \Val$ is the set of nondeterministic nodes,
$Q_D = ((\mathcal{L} \setminus \mathcal{L}_A) \times \Val) \cup \{\top\}$ is the set of deterministic nodes,
$\text{Init} = \{\bot\} \times I$ is the set of initial nodes,
$\text{Lab}_A = \{a_0, a_1\}$ is the set of action labels,
$\text{Lab}_P = \{\tau, 0, 1\}$ is the set of probabilistic labels,
\noindent and the relation $\rightarrow$ is defined as follows:
For every node $v = \tup{l, \sigma}$ of $\mathcal{M}_P$ and every edge $l \hookrightarrow l'$ of $P$
\begin{itemize}
  \item  if $\lab(l,l') = (x := \coin(p))$, then $v \xrightarrow {0, p} \tup{l', \sigma[x \mapsto 0]}$ and $v \xrightarrow  {1, 1-p} \tup{l', \sigma[x \mapsto 1]}$;
  \item  if $\lab(l,l') = (x := \nondet())$, then $v \xrightarrow {a_0} \tup{l', \sigma[x \mapsto 0]}$ and $v \xrightarrow {a_1}$
   \mbox{$\tup{l', \sigma[x \mapsto 1]}$};
  \item  if $\lab(l,l') = (x := e)$, then $v \xrightarrow {\tau, 1} \tup{l', \sigma[x \to e(\sigma)]}$,
    where $\sigma[x \to e(\sigma)]$ denotes the configuration obtained from~$\sigma$ by updating the value of~$x$ to
     the expression~$e$ evaluated under~$\sigma$;
  \item  if $\lab(l,l') = c$ for a conditional~$c$ satisfying~$\sigma$, then $v \xrightarrow {\tau, 1} \tup{l', \sigma}$.
\end{itemize}
\noindent For each node $v = \tup{\top, \sigma}$, $v \xrightarrow \tau \top$ and
  $\top \xrightarrow \tau \top$.
\qed

A program~$P = (\mathcal{L}, I, {\hookrightarrow}, \lab, \bot, \top)$ is called
\begin{itemize}
\item \emph{a.s.-terminating} if $\Pr^S_q[\{r \in \Runs(\mathcal{M}_P) \mid r \text{ reaches } \top \}] = 1$ for every strategy~$S$
 and every initial state~$q$ of $\mathcal{M}_P$;
 \item \emph{finite} if finitely many nodes are reachable from the initial nodes of~$\mathcal{M}_P$;
 \item \emph{weakly finite} if $P_b$ is finite for all $b \in I$,
   where $P_b$ is obtained from~$P$ by fixing~$b$ as the only initial node.
\end{itemize}
 \andreas{Bemerkung 3: ... ``We guess you mean that after fixing the value of the parameter
our programs always have a finite-state space. We'll clarify this.'' Das steht doch hier?}
\stefan{Ja.}

%

\emph{Assumption.} We assume in the following that programs to be analyzed are deterministic.
We consider nondeterministic programs only in Section~\ref{sub:nondet}.
\section{Patterns} \label{sec:patterns}

We introduce the notion of patterns for probabilistic programs.
A pattern restricts a probabilistic program by imposing particular sequences of coin toss outcomes on the program runs.
For the rest of the section we fix a probabilistic program $P = (\mathcal{L}, I, \mathord{\hookrightarrow}, \lab, \bot, \top)$
and its associated MDP $\mathcal{M}_P =  (Q_A, Q_P, \text{Init}, \mathord{\rightarrow}, \text{Lab}_A, \text{Lab}_P)$.


We write $C := \{0,1\}$ for the set of coin toss outcomes in the following.
A \emph{pattern} is a subset of $C^\omega$ of the form $C^* w_1 C^* w_2 C^* w_3 \ldots$, where $w_1, w_2 , \ldots \in \Sigma^*$.
We say the sequence $w_1, w_2, \ldots$ \emph{induces} the pattern.
Fixing an enumeration $x_1, x_2, \ldots$ of~$C^*$, we call the pattern induced by $x_1, x_2, \ldots$ the \emph{universal} pattern.
For a pattern~$\Phi$, a run $r \in \text{Runs}(\mathcal{M}_P)$ is \emph{$\Phi$-conforming}
if there is $v \in \Phi$ such that $\bar{r}|_C$ is a prefix of~$v$.
We call a pattern~$\Phi$ \emph{terminating (for~$P$)} if all $\Phi$-conforming runs terminate, i.e., reach~$\top$.
We show the following theorem:
\newcommand{\stmtthmcoinPattern}{
\begin{itemize}
\item[(1)]
 Let $\Phi$ be a pattern.
 The set of $\Phi$-conforming runs has probability~$1$.
 In particular, if $\Phi$ is terminating, then $P$ is a.s.-terminating.
\item[(2)]
 If $P$ is a.s.-terminating and weakly finite, then the universal pattern is terminating for~$P$.
\item[(3)]
 If $P$ is a.s.-terminating and finite with $n < \infty$ reachable nodes in~$\mathcal{M}_P$,
  then there exists a word $w \in C^*$ with $\abs{w} \in \mathcal{O}(n^2)$ such that $C^* w C^\omega$ is terminating for~$P$.
\end{itemize}
}
\begin{theorem} \label{th:coinPattern} \mbox{}
 \stmtthmcoinPattern
\end{theorem}

Part~(1) of Theorem~\ref{th:coinPattern} is the basis for the pattern approach.
It allows to ignore runs that are not $\Phi$-conforming, because they have probability~$0$.
Part~(2) states that the pattern approach is ``complete'' for a.s.-termination and weakly finite programs:
For any a.s.-terminating and weakly finite program there is a terminating pattern;
 moreover the universal pattern suffices.
Part~(3) refines part~(2) for finite programs: there is a short word such that $C^* w C^\omega$ is terminating.

\begin{proof}[of Theorem~\ref{th:coinPattern}] \\
Part~(1)~(Sketch):
We can show that the set of runs $r$ that visit infinitely many probabilistic nodes and do not have the form $C^* w_1 C^\omega$ is a null set.
This result can then easily be generalized to $C^* w_1 C^* w_2 \ldots C^* w_n C^\omega$. All runs conforming $\Phi$ can then be formed as a countable intersection of such run sets.

Part~(2):
Let $\sigma_1, \sigma_2, \ldots$ be a (countable or infinite) enumeration of the nodes in $I$.
With Part~(3) we obtain for each $i \geq 1$ a word $w_i$ such that $C^* w_i C^\omega$ is a terminating pattern for $P$,
 if the only starting node considered is $\sigma_i$.
By its definition, the universal pattern is a subset of $C^* w_i C^\omega$ for every $i \ge 1$, so it is also terminating.

Part~(3)~(Sketch): Since $P$ is a.s.-terminating, for every node $q$ there exists a coin toss sequence $w_q$, $\abs{w_q}\leq n$, with
the following property: a run that passes through $q$ and afterwards visits exactly the sequence $w_q$ of coin toss outcomes is terminating.
We build a sequence $w$ such that for every state $q$ every run that passes through $q$ and then visits exactly the sequence $w$ is terminating.
We start with $w = w_q$ for an arbitrary $q \not= \top$.
Then we pick a $q'\not=\top$ such that for $q'' \not= q$, runs starting in~$q''$ and visiting exactly the probabilistic label sequence $w$ lead to~$q'$.
We set $w = w_q w_{q'}$; after visiting $w$, all runs starting
from $q$ and $q'$ end in~$\top$. We iterate this until no more $q'$ can be found.
We stop after at most $n$ steps and obtain a sequence $w$ of length $\leq n^2$.
\qed
\end{proof}

\subsection{Nondeterministic Programs} \label{sub:nondet}

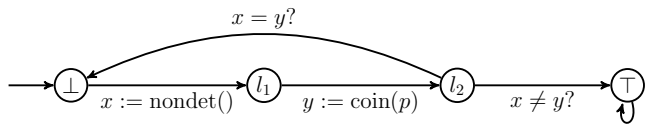
\begin{figure}[tbp]
\centering
 \tikzset{
   dim/.style={
           rounded corners,
           minimum height=2em,
           inner sep=2pt,
           text centered,
           },
    state/.style={
           rectangle,
           rounded corners,
           draw=black, very thick,
           minimum height=2em,
           inner sep=2pt,
           text centered,
           },
   act/.style={
           circle,
           draw=black, very thick,
           minimum height=2em,
           inner sep=2pt,
           text centered,
           },
  fail/.style={circle, double, draw=black, inner sep=2pt, minimum height=2em, very thick},
  nothing/.style={
           rectangle,
           rounded corners,
           minimum height=2em,
           inner sep=0pt,
           text centered,
           }
  }
\scalebox{0.6}{
\centering
\begin{tikzpicture}[->,>=stealth',scale=1,font=\Large]

 \node[dim, anchor=center] (S1)
 {$ $};

 \node[act] (L0) [right =of S1]
 {$\bot$};

 \node[act] (L1) [right =3.5cm of L0]
 {$l_1$};

 \node[act] (L2) [right =3.5cm of L1]
 {$l_2$};

 \node[act] (TOP) [right =3cm of L2]
 {$\top$};

 \path[very thick]
       (S1) edge[above] node [above] {$ $} (L0)
       (L0) edge[above] node [below] {$x := \nondet()$} (L1)
       (L1) edge[above] node [below] {$y := \coin(p)$} (L2)
       (L2) edge node [below] {$x \not= y?$} (TOP)
        (L2) edge[bend right=25] node [above] {$x = y?$} (L0)
       (TOP) edge[loop below] node [right] {} (TOP);
\end{tikzpicture}
}
\caption{Nondeterministic a.s.-terminating program without terminating pattern.}\label{fig:coinPatternCounterExample}
\end{figure}
For nondeterministic a.s.-terminating programs, there might not exist a terminating pattern, even if the program is finite.
Figure~\ref{fig:coinPatternCounterExample} shows an example.
%
Let $\Phi$ be a pattern and $c_1 c_2 c_3 \ldots \in \Phi$.
The run
\begin{equation*}
 \tup{\bot, \sigma_0}
 \xrightarrow {a_{c_1}}
 \tup{l_1, \sigma_1}
 \xrightarrow  {c_1}
 \tup{l_2, \sigma'_1}
 \xrightarrow \tau
 \tup{\bot, \sigma'_1}
 \xrightarrow  {a_{c_2}}
 \tup{l_1, \sigma_2}
 \xrightarrow  {c_2}
 \tup{l_2, \sigma'_2}
 \xrightarrow \tau
 \tup{\bot, \sigma'_2}
 \xrightarrow  {a_{c_3}}
 \ldots
\end{equation*}
in $\mathcal{M}_P$ is $\Phi$-conforming but nonterminating.

We show that the concept of patterns can be suitably generalized to nondeterministic programs,
 recovering a close analog of Theorem~\ref{th:coinPattern}.
Assume that the program is in a normal form where nondeterministic and probabilistic locations strictly alternate.
This is easily achieved by adding dummy assignments.
Writing $A := \{a_0, a_1\}$, every run $r \in \mathcal{M}_P$ satisfies $r|_{A \cup C} \in (A C)^\infty$.

A \emph{response} of length~$n$ encodes a mapping $A^n \to C^n$ in an ``interleaved'' fashion,
 e.g., $\{a_01, a_10\}$ is a response of length one, $\{a_00a_01, a_00a_11, a_10a_01, a_10a_11\}$ is a response of length two.
A \emph{response pattern} is a subset of $(A C)^\omega$ of the form $(A C)^* R_1 (A C)^* R_2 (A C)^* \ldots$, where $R_1, R_2, \ldots$ are responses.
If we now define the notions of \emph{universal} and \emph{terminating} response patterns analogously to the deterministic case,
 a theorem very much like Theorem~\ref{th:coinPattern} can be shown.
For instance, let $\Phi = (A C)^* \{a_01, a_10\} (A C)^\omega$.
Then every $\Phi$-conforming run of the program in Fig.~\ref{fig:coinPatternCounterExample} has the form
\begin{equation*}
 \tup{\bot, \sigma_0} \rightarrow \ldots \rightarrow q \xrightarrow {a_{i} } q' \xrightarrow {1-i} q'' \rightarrow \top  \rightarrow \ldots
 \qquad \text{for an $i \in \{0,1\}$.}
\end{equation*}
This implies that the program is a.s.-terminating (for all strategies).
\iftechrep{See Appendix~\ref{app:response}}{See~\cite{patternTechRep}} for the details.  
\section{Our Algorithm} \label{sec:computing}

In this section we aim at a procedure that, given a weakly finite program~$P$, proves that $P$ is a.s.-terminating by computing a terminating pattern.
This approach is justified by Theorem~\ref{th:coinPattern}~(1).
In fact, the proof of Theorem~\ref{th:coinPattern}~(3) constructs, for any finite a.s.-terminating program, a terminating pattern.
However, the construction operates on the Markov chain~$\mathcal{M}_P$, which is expensive to compute.
To avoid this, we would like to devise a procedure which operates on~$P$,
 utilizing (nonprobabilistic) verification tools, such as model checkers and termination provers.

Theorem~\ref{th:coinPattern}~(2) guarantees that, for any weakly finite a.s.-terminating program,
 the universal pattern is terminating.
This suggests the following method for proving a.s.-termination of~$P$:
 (i) replace in~$P$ all probabilistic assignments by nondeterministic ones
  and instrument the program so that all its runs are conforming to the universal pattern
   (this can be done as we describe in Section~\ref{sub:pattern-checker} below);
then (ii) check the resulting program for termination with a termination checker such as ARMC~\cite{armc}.
Although this approach is sound and complete (modulo the strength of the termination checker),
 it turns out to be useless in practice.
This is because the crucial loop invariants are extremely hard to catch for termination checkers.
Already the instrumentation that produces the enumeration of~$C^*$ requires a nontrivial procedure (such as a binary counter)
 whose loops are difficult to analyze.

Therefore we devise in the following another algorithm which tries to compute a terminating pattern $C^* w_1 C^* w_2 \ldots$
It operates on~$P$ and is ``refinement''-based.
Our algorithm uses a  ``pattern checker'' subroutine which takes a sequence $w_1, w_2, \ldots$, and checks (or attempts to check)
 whether the induced pattern is terminating.
If it is not, the pattern checker may return a \emph{lasso} as counterexample.
Formally, a lasso is a sequence
 \[
  \tup{l_1, \sigma_1} \to \tup{l_2, \sigma_2} \to \ldots \to \tup{l_m, \sigma_m} \to \ldots \to \tup{l_n, \sigma_n}
   \quad \text{with $\tup{l_n, \sigma_n} \to \tup{l_m, \sigma_m}$}
 \]
and $\tup{l_1, \sigma_1} \in \text{Init}$.
We call the sequence $\tup{l_m, \sigma_m} \to \ldots \to \tup{l_n, \sigma_n}$ the \emph{lasso loop} of the lasso.
Note that a lasso naturally induces a run in $\Runs(\mathcal{M}_P)$.
If $P$ is finite, pattern checkers can be made complete, i.e., they either prove the pattern terminating or return a lasso.

We present our pattern-finding algorithms for finite-state and weakly finite programs.
In Section~\ref{sub:pattern-checker} we describe how pattern-finding and pattern-checking
can be implemented using existing verification tools.

\subsubsection*{Finite Programs.}

First we assume that the given program~$P$ is finite.
The algorithm may take a \emph{base word} $s_0 \in C^*$ as input, which is set to $s_0 := \epsilon$ by default.
Then it runs the pattern checker on $C^* s_0 C^* s_0 \ldots$
If the pattern checker shows the pattern terminating, then, by Theorem~\ref{th:coinPattern}~(1), $P$ is a.s.-terminating.
Otherwise the pattern checker provides a lasso
 $\tup{l_1, \sigma_1} \to \ldots \to \tup{l_m, \sigma_m} \to \ldots \to \tup{l_n, \sigma_n}$.
Our algorithm extracts from the lasso loop a word $u_1 \in C^*$,
 which indicates a sequence of outcomes of the coin tosses in the lasso loop.
If $u_1 = \epsilon$, then the pattern checker has found a nonterminating run with only finitely many coin tosses,
 hence $P$ is not a.s.-terminating.
Otherwise (i.e., $u_1 \ne \epsilon$), let $s_1 \in C^*$ be a shortest word such that $s_0$ is a prefix of~$s_1$ and $s_1$ is not an infix of~$u_1^\omega$.
Our algorithm runs the pattern checker on $C^* s_1 C^* s_1 \ldots$
If the pattern checker shows the pattern terminating, then $P$ is a.s.-terminating.
Otherwise the pattern checker provides another lasso, from which our algorithm extracts a word~$u_2 \in C^*$ similarly as before.
If $u_2 = \epsilon$, then $P$ is not a.s.-terminating.
Otherwise, let $s_2 \in C^*$ be a shortest word such that $s_0$ is a prefix of~$s_2$ and $s_2$ is neither an infix of~$u_1^\omega$ nor an infix of~$u_2^\omega$.
Observe that the word $s_1$ is an infix of~$u_2^\omega$ by construction, hence $s_2 \ne s_1$.
Our algorithm runs the pattern checker on $C^* s_2 C^* s_2 \ldots$ and continues similarly,
 in each iteration eliminating all lassos so far discovered.

The algorithm is complete for finite and a.s.-terminating programs:

\newcommand{\stmtpropfinitealgo}{
Let $P$ be finite and a.s.-terminating.
Then the algorithm finds a shortest word~$w$ such that the pattern $C^* w C^* w \ldots$ is terminating, thus proving termination of~$P$.
}
\begin{proposition} \label{prop:finitealgo}
 \stmtpropfinitealgo
\end{proposition}

In each iteration the algorithm picks a word~$s_j$ that destroys all previously discovered lasso loops.
If the loops are small, then the word is short:
\newcommand{\stmtpropshortword}{
 We have $|s_j| \le |s_0| + 1 + \log_2 \left( |u_1| + \cdots + |u_j| \right)$.
}
\begin{proposition} \label{prop:shortword}
 \stmtpropshortword
\end{proposition}

The proofs for both propositions can be found in\iftechrep{Appendix~\ref{app:computingproofs}}{~\cite{patternTechRep}}.

\subsubsection*{Weakly Finite Programs.}
Let us now assume that $P$ is a.s.-terminating and weakly finite.
We modify our algorithm.
Let $b_1, b_2, \ldots$ be an enumeration of the set~$I$ of initial nodes.
Our algorithm first fixes~$b_1$ as the only initial node.
This leads to a finite program, so we can run the previously described algorithm,
 yielding a word~$w_1$
 such that $C^* w_1 C^* w_1 \ldots$ is terminating for the initial node~$b_1$.
Next our algorithm fixes~$b_2$ as the only initial node,
 and runs the previously described algorithm taking~$w_1$ as base word.
As before, this establishes a terminating pattern $C^* w_2 C^* w_2 \ldots$
By construction of~$w_2$, the word~$w_1$ is a prefix of~$w_2$,
 so the pattern $C^* w_1 C^* w_2 C^* w_2 \ldots$ is terminating for the initial nodes $\{b_1, b_2\}$.
Continuing in this way we obtain a sequence $w_1, w_2, \ldots$ such that
 $C^* w_1 C^* w_2 \ldots$ is terminating.
Our algorithm may not terminate, because it may keep computing $w_1, w_2, \ldots$.
However, we will illustrate that it is promising to compute the first few $w_i$ and then \emph{guess} an expression for general $w_i$.
For instance if $w_1 = 0$ and $w_2 = 0 0$, then one may guess $w_i = 0^i$.
We encode the guessed sequence $w_1, w_2, \ldots$ in a finite way and pass the obtained pattern $C^* w_1 C^* w_2 \ldots$ to a pattern checker,
 which may show the pattern terminating,
  establishing a.s.-termination of the weakly finite program~$P$.

%
%
%

\subsection{Implementing Pattern Checkers} \label{sub:pattern-checker}
\subsubsection*{Finite Programs.}
\label{sub:pattern-checker1}
We describe how to build a pattern checker for finite programs $P$ and patterns of the form $C^* w C^* w \ldots$
We employ a model checker for finite-state nonprobabilistic programs that can verify temporal properties:
Given as input a finite program and a Büchi automaton $\mathcal{A}$, the model checker returns a lasso if there is a program run
accepted by $\mathcal{A}$ (such runs are called ``counterexamples'' in classical terminology).
Otherwise it states that there is no counterexample.
For our case studies, we use the SPIN tool~\cite{spin}.

Given a finite probabilistic program $P$ and a pattern $\Phi = C^* w C^* w \ldots$,
we first transform $P$ into a nonprobabilistic program $P'$ as follows.
We introduce two fresh variables $c$ and $\text{term}$, with ranges $\{0,1,2\}$ and $\{0,1\}$, respectively,
 and add assignments \verb+term := 0+ and \verb+term := 1+ at the beginning and end of the program, respectively.
Then every location $l$ of $P$ with $label(l,l') = x := \coin(p)$ for a label $l'$ is replaced by a nondeterministic choice and an
if-statement as follows:
\begin{minipage}{20cm}
\begin{verbatim}
x := nondet();
if (x = 0)  c := 0; c := 2; else c := 1; c := 2; end if;
\end{verbatim}
\end{minipage}
In this way we can distinguish coin toss outcomes in a program trace by inspecting the assignments to~$c$.
Now we perform two checks on the nonprobabilistic program~$P'$:

First, we use SPIN to translate the LTL formula $G\, \neg \text{term} \land F G (c \not\in \{0,1\})$ into a Büchi automaton
 and check whether $P'$ has a run that satisfies this formula.
If there is indeed a lasso, our pattern checker reports it.
Observe that by the construction of the LTL formula the lasso encodes a nonterminating run in~$P$ that eventually stops visiting probabilistic locations.
So the lasso loop does not contain any coin tosses (and our algorithm will later correctly report that $P$ is not a.s.-terminating).
Otherwise, i.e., if no run satisfies the formula, we know that all nonterminating runs involve infinitely many coin tosses.
Then we perform a second query:
We construct a Büchi automaton $\mathcal{A}(w)$ that represents the set of infinite $\Phi$-conforming runs,
 see 
 Fig.~\ref{fig:buechi_automaton}.
We use SPIN to check whether $P'$ has run that is accepted by~$\mathcal{A}(w)$.
If yes, then there is an infinite $\Phi$-conforming run, and our pattern checker reports the lasso.
Otherwise, it reports that $\Phi$ is a terminating pattern.

\begin{figure}[tbz]
\centering
 \tikzset{
   dim/.style={
           rounded corners,
           minimum height=2em,
           inner sep=2pt,
           text centered,
           },
    state/.style={
           circle,
           rounded corners,
           draw=black, very thick,
           minimum height=2em,
           inner sep=2pt,
           text centered,
           },
    fstate/.style={
           circle,
           double,
           rounded corners,
           draw=black, very thick,
           minimum height=2em,
           inner sep=2pt,
           text centered,
           },
   act/.style={
           circle,
           draw=black, very thick,
           minimum height=2em,
           inner sep=2pt,
           text centered,
           },
  fail/.style={circle, double, draw=black, inner sep=2pt, minimum height=2em, very thick},
  nothing/.style={
           rectangle,
           rounded corners,
           minimum height=2em,
           inner sep=0pt,
           text centered,
           }
  }
\scalebox{0.65}{
\begin{tikzpicture}[->,>=stealth',scale=1,font=\Large]

 \node[dim, anchor=center] (S1)
 {$\,\,\,\,\,\,\,$};

 \node[state] (A1) [right =0.5cm of S1]
 {$\,\,\,\,\,\,\,$};

 \node[state] (A2) [right =2cm of A1]
 {$\,\,\,\,\,\,\,$};

 \node[state] (A3) [right =2cm of A2]
 {$\,\,\,\,\,\,\,$};

 \node[dim] (Z) [right =2cm of A3]
 {$\Huge{\ldots}$};

 \node[state] (A4) [right =2cm of Z]
 {$\,\,\,\,\,\,\,$};

 \node[fstate] (A5) [right =2cm of A4]
 {$\,\,\,\,\,\,\,$};

 \path[very thick]
       (S1) edge[above] node [below] {$ $} (A1)
       (A1) edge[above] node [below] {$ c = c_1$} (A2)
       (A2) edge[above] node [below] {$ c = c_2$} (A3)
       (A3) edge[above] node [below] {$ c = c_3$} (Z)
       (Z)  edge[above] node [below] {$ c = c_{n-1}$} (A4)
       (A4) edge[above] node [below] {$ c = c_n$} (A5)
       (A1) edge[loop below] node [below] {$\text{true}$} (A1)
       (A2) edge[loop below] node [below] {$c = 2$} (A2)
       (A3) edge[loop below] node [below] {$c = 2$} (A3)
       (A4) edge[loop below] node [below] {$c = 2$} (A4)
       (A5) edge[bend right=20] node [below] {$\text{true}$} (A1);
\end{tikzpicture}
}
\caption{Büchi automaton $\mathcal{A}(w)$, for $w = c_1c_2\ldots c_n \in C^*$. 
Note that the number of states in $\mathcal{A}(w)$ grows linearly in $\abs{w}$.}
\label{fig:buechi_automaton}
\end{figure}
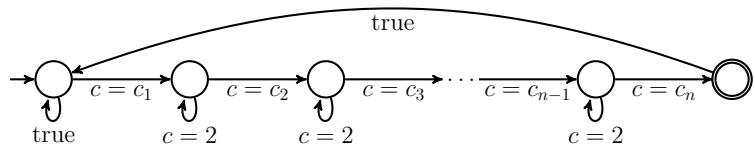

\subsubsection*{Weakly Finite Programs.}
\label{sub:pattern-checker2}

Recall that for weakly finite programs, the pattern checker needs to handle patterns of a more general form,
 namely $\Phi = C^* w_1 C^* w_2 \ldots$
Even simple patterns like $C^* 0 C^* 00 C^* 000 \ldots$ cannot be represented by a finite Büchi automaton.
Therefore we need a more involved instrumentation of the program to restrict its runs to $\Phi$-conforming ones.
Now our pattern checker employs a termination checker for infinite-state programs.
For our experiments we use ARMC.

Given a weakly finite program~$P$ and a pattern $\Phi = C^* w_1 C^* w_2 \ldots$, we transform~$P$ into a nonprobabilistic program $P^\Phi$ as follows.
We will use a command \verb+x := ?+, which nondeterministically assigns a nonnegative integer to $x$.
Further we assume that we can access the $k$-th letter of the $i$-th element of $(w_i)_{i \in \mathbb{N}}$ by $\text{w[i][k]}$,
and $\abs{w_i}$ by $\text{length(w[i])}$.
We add fresh variables \verb+ctr+, \verb+next+ and \verb+pos+,
 where \verb+ctr+ is initialized nondeterministically with any nonnegative integer
  and \verb+next+ and \verb+pos+ are both initialized with~1.
If a run~$r$ is $\Phi$-conforming, $\bar{r}|_C$ is a prefix of $v_1 w_1 v_2 w_2 v_3 w_3 \ldots$, with $v_i \in C^*$.
The variable \verb+ctr+ is used to ``guess'' the length of the words $v_i$;
 the individual letters in~$v_i$ are irrelevant.
We replace every command $c := \coin(p)$ by the code sequence given in Fig.~\ref{fig:armc_program_trans}.

The runs in the resulting program~$P^\Phi$ correspond exactly to the $\Phi$-conforming runs in~$P$.
Then $P^\Phi$ is given to the termination checker.
If it proves termination, we report ``$\Phi$ is a terminating pattern for $P$''.
Otherwise, the tool might either return a lasso, which our pattern checker reports, or give up on $P^\Phi$, in which case our pattern checker also has to give up.

In our experiments, a weakly finite program typically has an uninitialized integer variable~$N$ whose value is nondeterministically fixed in the beginning.
The pattern $C^* w_1 C^* \ldots C^* w_N C^\omega$ is then often terminating,
 which makes $\texttt{next} \le N$ an invariant in~$P^\Phi$.
The termination checker ARMC may benefit from this invariant, but may not be able to find it automatically
 (for reasons unknown to the authors).
We therefore enhanced ARMC to ``help itself'' by adding the invariant $\text{next} \le N$ to the program if ARMC's reachability mode can verify the invariant.

\begin{figure}[tbz]
\begin{minipage}{20cm}
\begin{verbatim}
x := nondet();
if (ctr <= 0)
  if (pos > length(w[next])) ctr := ?; pos := 1; next := next+1;
  else x := w[next][pos]; pos := pos+1;
else ctr := ctr-1;
\end{verbatim}
\end{minipage}%
\caption{Code transformation for coin tosses in weakly finite programs.}
\label{fig:armc_program_trans}
\end{figure}

\section{Experimental evaluation}
We apply our methods to several parameterized programs taken from the literature.\footnote{The sources can be found at
\url{http://www.model.in.tum.de/~gaiser/cav2012.html}.}

\begin{itemize}
\item\emph{firewire:} Fragment of FireWire's symmetry-breaking protocol, adapted from~\cite{mciver}
(a simpler version was used in the introduction). Roughly speaking, the number 100 of Fig.~\ref{fig:firewire}
is replaced by a parameter~$N$.
\item\emph{randomwalk:} A slightly different version of the finite-range, one-dimensional random walk used as second example in the introduction.
\item\emph{herman:} An abstraction of Herman's randomized algorithm for leader election used in~\cite{nakata}.
It can be seen as a more complicated finite random walk, with $N$ as the walk's length.
\item\emph{zeroconf:} A model of the Zeroconf protocol taken from~\cite{kattenbelt}.
The protocol assigns IP addresses in a network. The parameter $N$ is the number of probes sent after
choosing an IP address to check whether it is already in use.
\item\emph{brp:} A model adapted from~\cite{kattenbelt} that models the well-known
bounded retransmission protocol. The original version can be proven a.s.-terminating with the trivial pattern $C^\omega$; hence we
study an ``unbounded'' version, where arbitrarily many retransmissions are allowed.
The parameter $N$ is the length of the message that the sender must transmit to the receiver.
\end{itemize}

\subsubsection*{Proving a.s.-termination.} We prove a.s.-termination of the examples using SPIN~\cite{spin} to find patterns of
finite-state instances, and ARMC~\cite{armc} to prove termination of the nondeterministic programs
derived from the guessed pattern. All experiments were performed on an Intel\textsuperscript{\textcopyright} i7 machine with 8GB RAM.
The results are shown
in Fig.~\ref{tab:casestudies}.
The first two columns give the name of the example and its size. The next two columns show the words
$w_1,\ldots, w_4$ of the terminating patterns $C^* w_1 C^\omega, \ldots, C^* w_4 C^\omega$ computed for $N=1,2,3,4$
(see Theorem \ref{th:coinPattern}(3) and Section \ref{sub:pattern-checker1}),
and SPIN's runtime. The last two columns
give word $w_i$ in the guessed pattern $C^* w_1 C^* w_2 C^* w_3 \ldots$ (see Section \ref{sub:pattern-checker2}), and
ARMC's runtime. For instance, the entry $0(10)^i$
for {\it herman} indicates that the guessed pattern is $C^* 010 C^* 01010 C^* 0101010 \ldots$.

\begin{figure}[tz]
 \begin{center}
 \begin{tabular}{l|c|cccc|c|c|c}
 Name & \#loc  & \multicolumn{4}{c|}{Pattern words for}    & Time     & $i$-th word of & Time             \\
      &        & \multicolumn{4}{c|}{$N=1,2,3,4$}          & (SPIN)   & guessed pattern & (ARMC)              \\ \hline

 \emph{firewire}       &  19 & $010$ & $010$ & $010$ &  $010$ &  17 sec
& 010 & \textcolor{white}{00}1 min 36 sec   \\ \hline
 \emph{randomwalk}     &  16  & $\epsilon$ & $0^2$ &  $0^3$ &  $0^4$ & 23 sec &
$0^{i}$ & \textcolor{white}{00}1 min 22 sec \\ \hline
 \emph{herman}         &  36 & $010$ & $0(10)^2$ & $0(10)^3$ & $0(10)^4$ & 47 sec &
$0(10)^{i}$ & \textcolor{white}{00}7 min 43 sec\\ \hline
 \emph{zeroconf}       &  39 & $0^3$  & $0^4$ &  $0^5$ & $0^6$ & 20 sec &
$0^{i+2}$  & \textcolor{white}{0}26 min 16 sec  \\ \hline
 \emph{brp}       &  57 & $00$   & $00$  & $00$  & $00$ &  19 sec &
$00$ & \textcolor{white}{0}45 min 14 sec  \\
 \end{tabular}
 \end{center}
 \caption{Constructed patterns of the case studies and runtimes.}
 \label{tab:casestudies}
 \end{figure}

We derive two conclusions. First, a.s.-termination is proved
by very simple patterns: the general shape is easily guessed
from patterns for $N=1,2,3,4$, and the need for human ingenuity is virtually reduced to zero.
This speaks in favor of the Planner technique of \cite{DBLP:conf/fossacs/AronsPZ03}
and our extension to patterns, compared to other approaches using fairness and Hoare calculus \cite{PnueliProbVerification,McIverPGCL}.
Second, the runtime is dominated by the termination tool, not by the finite-state
checker. So the most direct way to improve the efficiency of our technique is to
produce faster termination checkers.

In the introduction we claimed that general purpose probabilistic model-checkers
perform poorly for a.s.-termination, since they are not geared towards this
problem. To supply some evidence for this, we tried
to prove a.s.-termination of the first four examples using
the CEGAR-based PASS model checker \cite{wachter,hermanns}. In all four cases the refinement loop did not terminate.\footnote{Other checkers, like PRISM, cannot be applied because they
only work for finite-state systems.}

\subsubsection*{Improving  lower bounds for reachability.} Consider a program of the form
\verb+if coin(0.8) {P1(); else P2()}; ERROR+ . Probabilistic model-checkers
compute lower and upper bounds for the probability of ERROR. Loosely speaking, lower bounds are computed
by adding the probabilities of terminating runs of $\text{P1}$ and~$\text{P2}$. However, since CEGAR-based checkers \cite{wachter,hermanns,kattenbelt,DBLP:conf/sas/EsparzaG11} work with abstractions of $\text{P1}$ and~$\text{P2}$, they may not be able to ascertain that paths
of the abstraction are concrete paths of the program, leading to poor lower bounds. Information on a.s.-termination
helps: if e.g.\ $\text{P1}$ terminates a.s., then we already have a lower bound of $0.8$.
We demonstrate this technique on two examples. The first one is the following modification of
{\it firewire}:
\begin{center}
\begin{minipage}{10cm}
\begin{verbatim}
N = 1000; k = 0; miss = 0;
while (k < N) {
  old_x = x; x = coin(0.5);
  if (x = old_x) k++ else if (k < 5) miss = 1
}
\end{verbatim}
\end{minipage}
\end{center}
For $i \in \{0,1\}$, let $p_i$ be the probability that the program terminates with \mbox{$\text{miss} = i$}.
After 20 refinement steps PASS returns upper bounds of $0.032$ for~$p_0$ and $0.969$ for $p_1$,
but a lower bound of $0$ for $p_1$, which stays~$0$ after $300$ iterations.
Our algorithm establishes that the loop a.s.-terminates, which implies $p_0 + p_1 = 1$,
and so after 20 iterations we already get $0.968 \le p_1 \le 0.969$.

We apply the same technique to estimate the probabilities $p_1, p_0$ that {\it zeroconf}
detects/does-not-detect an unused IP address. For $N = 100$, after 20 refinement steps PASS reports
an upper bound of $0.999$ for~$p_0$, but a lower bound of~$0$ for~$p_1$, which stays~$0$
for $80$ more iterations. With our technique after 20 iterations we get $0.958 \le p_1 \le 0.999$.

\section{Conclusions} \label{sec:conclusions}

We have presented an approach for automatically proving a.s.-termination of probabilistic programs.
Inspired by the Planner approach of \cite{DBLP:conf/fossacs/AronsPZ03}, we instrument a probabilistic program $P$
into a nondeterministic program $P'$ such that the runs of $P'$ correspond to a set of runs
of $P$ with probability~$1$. The instrumentation is fully automatic for finite-state programs, and
requires an extrapolation step for weakly finite programs. We automatically check termination of $P'$
profiting from new tools that were not available to \cite{DBLP:conf/fossacs/AronsPZ03}. While our approach
maintains the intuitive appeal of the Planner approach, it allows to prove completeness results.
Furthermore, while in \cite{DBLP:conf/fossacs/AronsPZ03} the design of the Planner was left to the verifier,
we have provided in our paper a CEGAR-like approach. In the case of parameterized programs, the approach
requires an extrapolation step, which however in our case studies proved to be straightforward.
Finally, we have also shown that our approach to improve the game-based CEGAR technique of
\cite{wachter,hermanns,kattenbelt} for computing upper and lower bounds for the probability of reaching a program location.
While this technique often provides very good upper bounds, the lower bounds
are not so satisfactory (often $0$), due to spurious nonterminating runs introduced by the abstraction.
Our approach allows to remove the effect of these runs.

In future work we plan to apply learning techniques to pattern generation,
thereby inferring probabilistic termination arguments for large program instances from small instances.

\subsubsection*{Acknowledgments.}
We thank several anonymous referees for helping us clarify certain aspects of the paper.
We also thank Corneliu Poppea and Andrey Rybalchenko for many discussions and their help with ARMC,
and Björn Wachter and Florian Zuleger for fruitful insights on quantitative probabilistic analysis and termination techniques.

\bibliographystyle{plain} 
\bibliography{literatur}

\newpage
\iftechrep{
\appendix
\section*{Appendix}

In Appendix~\ref{app:response} we give details on patterns for nondeterministic programs.
Appendix~\ref{app:proofs} contains additional preliminaries that are needed for the following appendices:
In~\ref{app:coinproofs} and~\ref{app:computingproofs} proofs for Sections~\ref{sec:patterns} respectively
Section~\ref{sec:computing} are given. Appendix~\ref{app:responseproofs} contains a proof for the theorem in Appendix~\ref{app:response}.

\section{Patterns for Nondeterministic Programs}
\label{app:response}

For general a.s.-terminating probabilistic programs, there might not exist a terminating pattern, even if
the program is finite, recall Figure~\ref{fig:coinPatternCounterExample}.

We therefore propose another pattern class that also takes nondeterministic decisions into account.
We fix an arbitrary probabilistic program~$P = (\mathcal{L}, I, \mathord{\hookrightarrow}, \lab, \bot, \top)$,
and its associated MDP $\mathcal{M}_P =  (Q_A, Q_P, \text{Init}, \mathord{\rightarrow}, \text{Lab}_A, \text{Lab}_P)$.
We assume that $P$ is in a special \emph{normal form}: Every nondeterministic location has a probabilistic location as its successor, every
probabilistic location has a nondeterministic location as its successor. It is easy to transform a program in normal form by adding
redundant probabilistic and nondeterministic locations such that the transformed program terminates if{}f the original one does.
For example, the program in Fig.~\ref{fig:coinPatternCounterExample} is in normal form.
If $P$ is in normal form, then every run $r \in \mathcal{M}_P$ is a prefix of a word in $(A C)^\infty$.

We write $A := \{a_0, a_1\}$ and $G := \{a_0, a_1\} \cup C$.
A set $W \subseteq (A C)^*$ is called a \emph{response} of length~$n \geq 0$ if
(i) every $w \in W$  has length $2n$,
(ii) for $w_1, w_2 \in W$ with $w_1 \not= w_2$, $w_1|_A \not= w_2|_A$ holds, and (iii) $W$ contains exactly $2^n$ elements.
We denote by $\Resp(n)$ the set of responses of length~$n$, and set $\Resp := \bigcup_{n \in \mathbb{N}}{\Resp(n)}$.
Intuitively, a response $R$ of length~$n$ contains for every sequence of nondeterministic actions of length $n$ a sequence of
coin toss outcomes of length $n$ (interleaved in one word of $R$). For example,
$\{a_01, a_10\}$ is a response of length one, $\{a_0 0a_0 1, a_0 0 a_1 1, a_1 0 a_0 1, a_1 0a_1 1\}$ is a response
of length two.

A \emph{response pattern} is a subset of $(A C)^\omega$ of the form $(A C)^* R_1 (A C)^* R_2 (A C)^* \ldots$, where $R_1, R_2, \ldots$ are responses.
We say $R_1, R_2, \ldots$ induces the response pattern.

As in the deterministic case,
fixing an enumeration $R_1, R_2, \ldots$ of all responses, we call the pattern $(A C)^* R_1 (A C)^*R_2(A C)^*$ a \emph{universal response pattern}.
For a response pattern~$\Phi$, a run $r \in \Runs(\mathcal{M}_P)$ is $\Phi$-conforming if there is $v \in \Phi$ such that $\bar{r}|_G$ is a prefix of $v$.
We call a response pattern $\Phi$ \emph{terminating} if all $\Phi$-conforming runs terminate.

Analogously to the deterministic case,
we show the following theorem in Appendix~\ref{app:responseproofs}:
\newcommand{\stmtthmresponsePattern}{
Let $P$ be a probabilistic program in normal form.
\begin{itemize}
\item[(1)]
 Let $\Phi$ be a response pattern.
 The set of $\Phi$-conforming runs has probability~$1$ for every strategy $S$ for $\mathcal{M}_P$.
 In particular, if $P$ has a terminating pattern, then $P$ is a.s.-terminating.
\item[(2)]
 If $P$ is a.s.-terminating and weakly finite, then the universal pattern is terminating for~$P$.
\item[(3)]
 If $P$ is a.s.-terminating and finite with $n < \infty$ reachable nodes in~$\mathcal{M}_P$,
  then there exists a response $R$ of length in $\mathcal{O}(n^2)$ such that $(A C)^* R (A C)^\omega$ is terminating for~$P$.
\end{itemize}
}
\begin{theorem} \label{th:responsePattern}
 \stmtthmresponsePattern
\end{theorem}
%

\section{Proofs}  \label{app:proofs}

\subsection*{Preliminaries} \label{app:prelim}
Let~$P = (\mathcal{L}, I, {\hookrightarrow}, \lab, \bot, \top)$, and let $\mathcal{M}_P$ be its corresponding
MDP with $\mathcal{M}_P = (Q_A, Q_D, \text{Init}, \rightarrow, \text{Lab}_A, \text{Lab}_P)$. 

The probability $\Pr[q_1 \xrightarrow l q_2]$ of a transition $q_1 \xrightarrow {l,p} q_2$ is equal to~$p$.
We define a probability measure on the set of runs of a Markov chain $\mathcal{M}$ in the usual way (see e.g.~\cite{principlesMC}).
The \emph{cylinder set} $\cyl{\pi}$ of a path~$\pi$ is the set of runs having $\pi$ as prefix.
The probability of $\cyl{\pi}$ for a path $\pi$ starting at $q$, denoted by $\Pr_{q}[\cyl{\pi}]$, is $1$ if $\pi = q$,
and otherwise the product of the probabilities of the transitions of $\pi$.
There is a unique extension of $\Pr_{q}$ to a probability measure
over the smallest $\sigma$-algebra $\mathfrak{S}^{\mathcal{M}}$
containing all cylinder sets starting at $q$. We denote it by $\Pr_{q}$. We say a set of runs is $\emph{measurable}$ if it
is contained in $\mathfrak{S}^{\mathcal{M}}$.

For general MDPs, we can only define a probability measure after resolving the nondeterminism.
A \emph{strategy} for an MDP~$\mathcal{M}$ is a function $S$ that maps the empty path
$\epsilon$ to an initial node $q_0$, and every path
$q_0 \xrightarrow {l_0} q_1 \cdots q_{n-1} \xrightarrow {l_{n-1}} q_n \in \Paths(\mathcal{M})$
 ending at an action node to a probability distribution over the outgoing labels of $q$, i.e.,
over the labels $l$ such that $q_n \xrightarrow l q$ for some node $q$. Given a strategy
$S$, we define the Markov chain $\mathcal{M}[S]$
as usual (see \cite{principlesMC} for a formal definition): the nodes of $\mathcal{M}[S]$
are the paths of $\mathcal{M}$ whose cylinders have nonzero probability, the transitions
are defined to match the definition of the nodes, and the transition probabilities are
assigned according to $S$. For every node $q$ of $\mathcal{M}$ we define a probability measure $\Pr_q^S$ over $\mathfrak{S}^{\mathcal{M}}$,
that assigns to a cylinder $\cyl{\pi}$ the probability
of $\cyl{\pi'}$ in the Markov chain $\mathcal{M}[S]$,
 where $\pi'$ is the unique path of $\mathcal{M}[S]$ starting at the node $q$ (notice that $q$ is also a path, and so a node of~$\mathcal{M}[S]$)
 and ending at the node $\pi$. We write $\Pr^S[\cdot]$ for $\Pr^S_{q_0}[\cdot]$.

\subsection{Proofs of Section~\ref{sec:patterns}} \label{app:coinproofs}


In this section we complete the proof of
\begin{qtheorem}{\ref{th:coinPattern}} \mbox{}
Let $P$ be a probabilistic program that is deterministic.
 \stmtthmcoinPattern
\end{qtheorem}

\begin{proof} $ $\\
It remains to prove parts (1) and (3).
Let~$P = (\mathcal{L}, I, {\hookrightarrow}, \lab, \bot, \top)$. Its corresponding
MDP $\mathcal{M}_P$ is a Markov chain $\mathcal{M}_P =  (Q, q_I, \mathord{\rightarrow}, \text{Lab}_P)$.
We write $Pr[\cdot]$ instead of $Pr_{q_I}[\cdot]$.
Let $\Phi = C^*w_1C^*w_2C^* \ldots$; the set of runs conforming to $\Phi$ are denoted by $\Runs(\Phi)$.



\subsubsection{Proof of Part (1):} $ $ \\
We first prove that $\Runs(\Phi)$ is measurable.
Let \emph{$I_C \in \mathfrak{S}^{\mathcal{M}_P}$} be the set of runs $r$ with $\bar{r}|_C \in C^\omega$.
For  $w_1, w_2, \ldots w_i$, $i \geq 1$, we define the set
\emph{$S(w_1, w_2, \ldots, w_i)$} by
\begin{equation*}
 S(w_1, w_2, \ldots, w_i) := \{ r \in \Runs(\mathcal{M}_P) \mid \bar{r}|_C \in C^* w_1 C^* w_2 C^* \ldots C^* w_i C^\omega \}.
\end{equation*}
$S(w_1, w_2, \ldots, w_i)$ is measurable:
Let $\text{NC}(i) \in \mathfrak{S}^{\mathcal{M}_P}$ be the set of all runs $r$ with the $i$-th label of $r$ not in $C$, and
$F(i, c)\in \mathfrak{S}^{\mathcal{M}_P}$ the set of runs that have $c$ as $i$-th label. Set
\begin{equation*}
 G(b^-,b^+, c_1 \ldots c_k) =  \bigcup_{b^- \leq a_1 < \ldots < a_k < b^+} \Big(
 \bigcap_{\substack{l > a_1 \\ \wedge l \not \in \{a_1, \ldots, a_k\}}}{\text{NC}(l)} \cap \bigcap_{1 \leq j \leq k} {F(a_j, c_j)} \Big) \in \mathfrak{S}^{\mathcal{M}_P}
\end{equation*}
for $\{b^-,b^+,k\} \subseteq \mathbb{N}$ and $c_1 \ldots c_k \in C^k$. $S(w_1, w_2, \ldots, w_n)$ can be written as
\begin{equation*}
I_C \cap  \bigcup_{0 \leq b^-_1 < b^+_1 < b^-_2 < \ldots < b^-_{n} < b^+_n} {G(b^-_1, b^+_1, w_1) \cap \ldots \cap G(b^-_{n}, b^+_n, w_n)} \in \mathfrak{S}^{\mathcal{M}_P}.
\end{equation*}
Since
\begin{equation*}
 \Runs(\Phi) = \Big( \Runs(\mathcal{M}_P) \setminus I_C\Big) \cup \bigcap_{i\geq 0}{S(w_1, w_2, \ldots, w_i)}  \in \mathfrak{S}^{\mathcal{M}_P}
\end{equation*}
we conclude that $\Runs(\Phi)$ is also measurable.

Next we show that $Pr[\Runs(\Phi)] = 1$.

For every prefix $w_1, w_2, w_3, \ldots, w_i$ of $(w_i)_{i \in \mathbb{N}}$,
$\Pr[S(w_1, \ldots, w_i)] =  \Pr[I_C]$ holds, i.e.,
the set of runs that visit probabilistic nodes infinitely often, but
are \emph{not} $C^*w_1C^*w_2C^*\ldots C^*w_iC^\omega$-conforming, have probability zero.

For proving this we write $w = w_1w_2\ldots w_i$. Let $n = \abs{w}$.
$S(w) \subseteq S(w_1, w_2, \ldots, w_i)$ holds for all $i$.
It suffices to show that $\Pr[S(w)] = \Pr[I_C]$, since this implies with $I_C \supseteq S(w_1, \ldots, w_i)$ that $\Pr[I_C] = \Pr[S(w_1, w_2, \ldots, w_i)]$.

Let $V(j)$ be the set of runs that visit a probabilistic node at least $j$ times,
and let
\begin{equation*}
B(j) =  V(j\cdot n) \cap (\Runs(\mathcal{M}_P) \setminus S(w))
\end{equation*}
be the set of runs $r$ that visit a probabilistic node at least $j\cdot n$ times, and $w$ is no substring of $\bar{r}|_C$.

Since there are only finitely many probabilistic locations in $P$, there exists a minimal probability $p_{\min} > 0$ such that for every
transition $q \xrightarrow {c,p'} q'$, $c \in \{0, 1\}$, $p' \geq p_{\min}$ holds.
We write $\text{NV}(w)$ (``not visited'') for the set of runs $r$ such that $\bar{r}|_C$ does not start with $w$.
Now
\begin{align*}
 & Pr[B(1)] & \\
  &\leq Pr[\text{NV}(w) \mid V(n)] \cdot Pr[V(n)]& \\
  &\leq (1-p_{\min}^n) \cdot Pr[V(j)] & \\
  &\leq (1-p_{\min}^n),
\end{align*}
i.e.,
after visiting probabilistic nodes at least $n$ times, the probability $p$ of \emph{not} seeing the sequence $w$ is at most $(1-p_{min}^n) < 1$.
With a simple inductive argument we obtain $Pr[B(j)] \le (1-p_{\min}^n)^j$.
It holds that $B(j) \supseteq B(j+1)$ for all $j$. Then
\begin{equation}
\label{eq:pr-b-is-zero}
Pr[\bigcap_{j \geq 1}{B(j)}] =  \lim_{j \rightarrow \infty}{Pr[B(j)]} \leq \lim_{j \rightarrow \infty}{(1-p_{min}^n)^j} = 0.
\end{equation}
We can write $S(w) = I_C \setminus \bigcap_{j \geq 0}{B(j)}$.
Hence
\begin{align*}
 Pr[S(w_1w_2 \ldots w_i)] &= Pr[I_C \setminus \bigcap_{j \geq 1}{B(j)}] &  \text{(Def. of $B(\cdot)$)}\\
 &= Pr[I_C] - Pr[\bigcap_{j \geq 1}{B(j) \cap I_C}] &\\
 &= Pr[I_C] & \text{(Eq.~\ref{eq:pr-b-is-zero}).}
\end{align*}

Now $Pr[I_C \setminus S(w_1, \ldots, w_i)] = Pr[I_C] - Pr[S(w_1, \ldots, w_i)] = 0$.
We can write
\begin{equation*}
 I_C \setminus \bigcap_{i\geq1}{S(w_1, \ldots, w_i)} = I_C \cap \bigcup_{i \geq 1}{I_C \setminus S(w_1, \ldots, w_i)}.
\end{equation*}
For every $i \geq 1$, $I_C \setminus S(w_1, \ldots, w_i)$ is a null set, thus the countable union
$\bigcup_{i \geq 1}{I_C \setminus S(w_1, \ldots, w_i)}$ is also a null set (*).

We conclude:
\begin{align*}
  & Pr[\Runs(\Phi)] & \\
  &= Pr[\Runs(\mathcal{M}_P) \setminus I_C] + Pr[\bigcap_{i\geq 0}{S(w_1, \ldots, w_i)} ] & \\
  &= Pr[\Runs(\mathcal{M}_P) \setminus I_C] + Pr[I_C] & \text{ (*)}\\
  &= 1.
\end{align*}

\subsubsection{Proof of Part (3):} $ $ \\

We say that $q \in Q$ \emph{ends up in} $q' \in Q$ \emph{following} $w = c_1 c_2 \ldots c_m \in C^*$ if
\begin{equation*}
 q \xrightarrow {\tau^* c_1 \tau^* c_2 \tau^* \ldots \tau^* c_m  \tau^*} q',
\end{equation*}
and $q'$ is probabilistic or $\top$. Note that $q'$ is unique if it exists, since $P$ is deterministic.

For every reachable node $q$ and every sequence $w \in C^*$ holds that either:
(i) $q$ ends up in a node following $w$, or (ii) $q$ ends up in $\top$
following a proper prefix of $w$. Otherwise there exists a reachable node $q' \not= \top$ from which no probabilistic location or $\top$ is reachable any more, which contradicts
that $P$ is a.s.-terminating.

For every node $q \in Q$, there exists a $w_q \in C^*$ such that
$q$ ends up in $\top$ following $w_q$, again due to the a.s.-termination property of $P$. We can choose $w_q$ such
that $\abs{w_q} < n$ by removing cycles.

We construct a sequence $w^{(0)},w^{(1)},\ldots, w^{(m)} $ using the following algorithm.
Set $w^{(0)} := \epsilon$ and $i := 0$.
\begin{enumerate}
\item Pick a $q'_i \in Q$ that does end up in a node $q_i \not= \top$ following $w^{(i-1)}$. If no such $q_i$ exists set $w := w^{(i-1)}$ and terminate.
\item Set $w^{(i)} := w^{(i-1)} w_{q_i}$. Set $i := i+1$ and go to 1).
\end{enumerate}

The node sets $Q^{(i)}$ consist of $\top$ and all nodes a state $q \in Q$ might end up after following $w^{(i)}$.
$Q^{(i)}$ contains at most $n-i$
nodes for every $i \geq 0$.
This is certainly true for $i = 0$.
In the $i$-th iteration the chosen $q'_i$ ends up in $q_i \in Q^{(i-1)}$
after following $w^{(i-1)}$. After following $w_{q_i}$, $q_i$ ends up in $\top$. Thus $q_i$ ends up in $\top$ after $w^{(i)}$. This implies that
$\abs{Q^{(i)}} < \abs{Q^{(i-1)}}$, since every node can end up in at most one node after following a nonempty coin sequence, and note that
every $Q^{(i)}$ contains $\top$.

After at most $n-1$ iterations, $\abs{Q^{(i)}} \leq 1$, and the algorithm terminates. Hence $\abs{w} \leq (n-1)\cdot \max_{q\in Q}{\abs{w_q}} \leq (n-1)^2$.
Every node of $Q$ ends up in $\top$ after following a prefix of $w^{(i)}$. If it ended up in another node $\hat{q}$, the algorithm would have
performed another iteration, making $w$ longer, if there were no node $q'$ such that $q$ ends up in $q'$, a prefix of $w$ must have led it to $\top$ before.

We can conclude that every run $r$ for which $\bar{r}|_C$ is a prefix of a word $C^* w C^\omega$ is terminating, and thus $C^* w C^\omega$ is a terminating pattern.
$ $
\qed
\end{proof}

%
%
%

\subsection{Proofs of Section~\ref{app:computingproofs}}
 \label{app:computingproofs}

\begin{qproposition}{\ref{prop:finitealgo}}
 \stmtpropfinitealgo
\end{qproposition}

\begin{proof}
Recall from the proof of Theorem~\ref{th:coinPattern}~(3) that there is a fixed word $z \in C^*$
 which leads from an arbitrary node in $\Runs(\mathcal{M}_P)$ to termination.
In particular, $z$ is never an infix of $u_i^\omega$ for any~$i$.
It follows that $s_0 z$ is never an infix of $u_i^\omega$ for any~$i$.
Assume for a contradiction that our algorithm does not succeed in proving termination.
Since the $s_i$ are all pairwise different, our algorithm eventually chooses $s_j := s_0 z$ for some $j \in \mathbb{N}$.
By the definition of~$z$ the pattern $C^* z C^* z \ldots$ is terminating,
 hence so is $C^* s_j C^* s_j \ldots$
It follows that the pattern checker shows $P_{j+1}$ terminating, which is a contradiction.
\qed
\end{proof}

\begin{qproposition}{\ref{prop:shortword}}
 \stmtpropshortword
\end{qproposition}
\begin{proof}
If a word~$w$ is not an infix of any of the words $u_1^\omega$, \ldots, $u_j^\omega$,
 then neither is $s_0 w$.
Hence it suffices to construct such a word~$w$ with $|w| \le 1 + \log_2 K$, where $K := |u_1| + \cdots + |u_j|$.
Let $p_1, \ldots, p_{K}$ be an enumeration of all suffixes of the words $u_1^\omega, \ldots, u_j^\omega$.
For any word~$w$, denote by $S(w) \subseteq \{p_1, \ldots, p_K\}$ the set of words $p \in \{p_1, \ldots, p_K\}$
 such that $w$ is a prefix of~$p$.
It suffices to construct~$w$ such that $|w| \le 1 + \log_2 K$ and $S(w) = \emptyset$.
We construct~$w$ iteratively.
Let $w_0 := \epsilon$.
In each iteration~$i$, choose $w_{i+1} := w_{i} c$ with $c \in \{0, 1\}$
 so that $|S(w_{i} c)|$ is minimized.
Observe that $|S(w_{i+1})| \le |S(w_{i})| / 2$, as all words in~$S(w_{i})$ start with either $w_{i} 0$ or~$w_{i} 1$.
It follows that $S\left(w_{1 + \lfloor \log_2 K \rfloor}\right) = \emptyset$.
\qed
\end{proof}

\subsection{Proofs of Appendix~\ref{app:response}}
 \label{app:responseproofs}

In this section we prove

\begin{qtheorem}{\ref{th:responsePattern}} \mbox{}
 \stmtthmresponsePattern
\end{qtheorem}

\begin{proof} $ $ \\

Let~$P = (\mathcal{L}, I, {\hookrightarrow}, \lab, \bot, \top)$.
The MDP corresponding to~$P$ is denoted by $\mathcal{M}_P = (Q_A, Q_D, \text{Init}, \rightarrow, \text{Lab}_A, \text{Lab}_P)$.
Let $\Phi = (A C)^* R_1 (A C)^* R_2 \ldots$, with $R_i$ a response for all $i \geq 1$.
We call the set of $\Phi$-corresponding runs $\Runs(\Phi)$.
For responses $R_1, R_2$ of length $n_1$ and $n_2$, respectively, and a word $w \in (A C)^+$, we set
$w \circ R_1 := \{ w r \mid r \in R_1 \} \text{ and } R_1 \circ R_2 := \{ r R_2 \mid r \in R_1\}$.
$R_1 \circ R_2$ is a response of length $n_1+n_2$.
We set $G := A \cup C$.
Recall that, since $P$ is in normal form, for every run $r$ in $\mathcal{M}_P$, $\bar{r}|_G$ is a prefix of a word
in $(AC)^\omega$.

\subsubsection{Proof of Part (1):}

We first prove that $\Runs(\Phi)$ is measurable.
Let \emph{$I_G \in \mathfrak{S}^{\mathcal{M}_P}$} be the set of runs $r$ with $\bar{r}|_G \in G^\omega$.
For  $R_1, R_2, \ldots R_i$, $i \geq 1$, we define the set
\emph{$S(R_1, R_2, \ldots, R_i)$} by
\begin{equation*}
 S(R_1, R_2, \ldots, R_i) := \{ r \in \Runs(\mathcal{M}_P) \mid \bar{r}|_G \in G^* R_1 C^* R_2 G^* \ldots G^* R_i G^\omega \}.
\end{equation*}
$S(R_1, R_2, \ldots, R_i)$ is measurable:
The set of runs $r$ such that $\bar{r}|_M \in G^* w_1 G^* w_2 G^* \ldots G^* w_i G^\omega$ for $(w_1, \ldots, w_i) \in R_1 \times \ldots \times R_i$ is measurable,
which can be proved by an easy variation of the first part of the proof of Theorem~\ref{th:coinPattern}.
$S(R_1, R_2, \ldots, R_i)$ is the finite union of all these run sets and thus is measurable.
Again, since
\begin{equation*}
 \Runs(\Phi) = \Big( \Runs(\mathcal{M}_P) \setminus I_G\Big) \cup \bigcap_{i\geq 0}{S(R_1, R_2, \ldots, R_i)}  \in \mathfrak{S}^{\mathcal{M}_P}
\end{equation*}
we conclude that $\Runs(\Phi)$ is  measurable.

Let $S$ be a strategy for $\mathcal{M}_P$.
We show that $Pr^S[\Runs(\Phi)] = 1$, again reusing ideas from the proof of Theoren~\ref{th:coinPattern}.

For every prefix $R_1, R_2, R_3, \ldots, R_i$ of $(R_i)_{i \in \mathbb{N}}$,
we show that $\Pr[S(R_1, \ldots, R_i)] =  \Pr[I_G]$ holds, i.e.,
the set of runs that visit probabilistic nodes infinitely often, but
are \emph{not} conforming to  $(AC)^*R_1(AC)^*\ldots (AC)^*R_i(AC)^\omega$
have probability zero.

For proving this we write $R = R_1\circ R_2 \circ \ldots R_i$. Let $n$ be the length of $R$.
$S(R) \subseteq S(R_1, R_2, \ldots, R_i)$ holds for all $i$.
Again it suffices to show that $\Pr[S(R)] = \Pr[I_G]$,
since this implies with $I_G \supseteq S(R_1, \ldots, R_i)$ that $\Pr[I_G] = \Pr[S(R_1, R_2, \ldots, R_i)]$.

We reuse the definition of the sets of runs
$V(j)$ that visit a probabilistic node at least $j$ times,
and set
\begin{equation*}
B(j) =  V(j\cdot n) \cap (\Runs(\mathcal{M}_P) \setminus S(R))
\end{equation*}
the set of runs $r$ that visit a probabilistic node at least $j\cdot n$ times, and \emph{no} $w \in R$ is a substring of $\bar{r}|_G$.
There exists a minimal probability $p_{min} > 0$ such that for every
transition $q \xrightarrow {c,p'} q'$ in $\mathcal{M}_P[S]$, $c \in \{0, 1\}$, $p' \geq p_{\min}$ holds. Note that this in general only holds for probabilistic
transitions labeled by $\{0,1\}$ in $\mathcal{M}_P[S]$.
For $x \in A^*$ we write $SC(x)$ (``strategy choice'') for the set of runs $r$ such that $\bar{r}|_A$ starts with $x$.
For $x \not= x'$ with $x,x' \in A^*$ having the same
length,
\begin{equation}
 \label{eq:SCdisjunct}
  SC(x) \cap SC(x') = \emptyset.
\end{equation}
Let $\text{NV}(w)$ (``not visited'') be again the set of runs $r$ such that $\bar{r}|_C$ does not start with $w \in C^*$.
With this we get
\begin{align*}
 & Pr^S[B(1)] & \\
  &\leq \sum_{w \in R} Pr^S[\text{NV}(\bar{w}|_C) \mid \text{SC}(\bar{w}|_A) \cap V(n)] \cdot Pr^S[\text{SC}(\bar{w}|_A) \cap V(n)]& \\
  &\leq \sum_{w \in R} (1-p_{\min}^n) \cdot Pr^S[SC(\bar{w}|_A) \cap V(j)] & \text{(Eq. \ref{eq:SCdisjunct})}\\
  &\leq (1-p_{\min}^n).
\end{align*}
Again we can see that after visiting probabilistic nodes at least $n$ times, the probability of \emph{not} seeing at least \emph{one} of the $w \in R$ is at most $(1-p_{min}^n) < 1$.
In $B(j)$, we repeat this
experiment at least $j$ times; by a simple inductive argument we get again $Pr^S[B(j)] \le (1-p_{min}^n)^j$.
Now we proceed exactly as in the proof of Theorem~\ref{th:coinPattern}, substituting $Pr[\cdot]$ by $Pr^S[\cdot]$, $I_C$ by $I_G$, and $S(w_1, \ldots w_i)$ by $S(R_1, \ldots, R_i)$,
and obtain $Pr^S[I_G] = Pr^S[\bigcap_{i \geq 0}{S(R_1, \ldots, R_i)}]$. We conclude
\begin{align*}
  & Pr^S[\Runs(\Phi)] & \\
  &= Pr^S[\Runs(\mathcal{M}_P) \setminus I_G] + Pr^S[\bigcap_{i\geq 0}{S(R_1, \ldots, R_i)} ] & \\
  &= Pr^S[\Runs(\mathcal{M}_P) \setminus I_G] + Pr^S[I_G] & \\
  &= 1.
\end{align*}

\subsubsection{Proof of Part (2):}

The proof proceeds analogously to the one of Theorem~\ref{th:coinPattern}, part (2):
Let $\sigma_1, \sigma_2, \ldots$ be a (countable or infinite) enumeration of the nodes in $I$.
With Part~(3) we obtain for each $i \geq 1$ a response $R_i$ such that $(AC)^* R_i (AC)^\omega$ is a terminating pattern for $P$,
if the only starting node considered is $\sigma_i$.
By its definition, the universal pattern is a subset of $(AC)^* R_i (AC)^\omega$ for every $i \ge 1$, so it is also terminating.

\subsubsection{Proof of Part (3):}

We reintroduce several notations from the proof of Theorem~\ref{th:coinPattern} and generalize them to accomodate nondeterminism.
We say now that $q \in Q$ \emph{ends up in} $q' \in Q$ \emph{following} $w = x_1 x_2 \ldots x_m \in G^*$ if
\begin{equation*}
 q \xrightarrow {\tau^* x_1 \tau^* x_2 \tau^* \ldots \tau^* x_m  \tau^*} q',
\end{equation*}
and $q'$ is probabilistic, nondeterministic, or $\top$. Again, if such a $q'$ exist, it is unique, since all transition choices are resolved.

For every reachable node $q \in Q_A$ and every sequence $w \in (AC)^*$ holds that either:
(i) $q$ ends up in a node following $w$, or (ii) $q$ ends up in $\top$
following a prefix of $w$. Otherwise there exists a node $q'$ from which no probabilistic or nondeterministic location or $\top$ is reachable any more, which contradicts
that $P$ is a.s.-terminating (note that there always exists a strategy that is able to cause $q_0$ ending up in $q'$ with nonzero probability, using
the nondeterministic choices given in $w$, see also below).

We show that for every node $q \in Q_A$ and every sequence $s_1 \ldots s_n \in A^n$
there exists a $c_1 c_2 \ldots c_n$ such that
$q$ ends up in $\top$ following a prefix of $s_1 c_1 \ldots s_n c_n$.
Assume for the sake of contradiction
that there exists $q \in Q_A$ and a sequence  $s_1 \ldots s_n \in A^n$ such that no
$c_1 \ldots c_n$ exists with the property described above.
We will construct a strategy $S$ such that (i) reaching $q$ has probability $> 0$,
(ii) every run reaching $q$ will never reach $\top$.
The probability of reaching $\top$ is then smaller than 1, contradicting the assumption that $P$ is a. s. terminating.
Recall that nodes of $\mathcal{M}_P[S]$ are paths in $\mathcal{M}_P$.
Since $q$ is reachable in $\mathcal{M}_P$, there exists a cycle-free path $\pi$ from the initial node $q_0$ to $q$.
For all proper path prefixes of $\pi$ ending in a nondeterministic node, $S$ selects the corresponding choices contained in $\pi$ with probability 1, and
thus we reach $q$ with probability $> 0$.
For (ii), let $\pi$ be a path having the form $\pi = \pi' \rightarrow q_1 \xrightarrow {l_1} q_2 \xrightarrow {l_2} \ldots \xrightarrow {l_m} q_m$, with $m \geq 1$ and $q_1 = q$, such that $\pi'$ does not contain $q$.
We define $S(\pi)$ as follows:
Let $\pi_{r}$ be the path obtained from $q_1 \xrightarrow {l_1} q_2 \xrightarrow {l_2} \ldots \xrightarrow {l_m} q_m$ by removing all possible cycles.
$\pi_r$ then contains $k < n$ nondeterministic
nodes (there are only $n$ nodes in total). Set $S(\pi)(s_k) = 1$. Then there is no path starting from a reachable node $\pi' \rightarrow q$ in $\mathcal{M}_P[S]$ that reaches $\top$ (more exactly, that reaches a node $\pi''\rightarrow \top$),
contradicting the assumption that $P$ is a.s.-terminating.

We now select a $c_1 \ldots c_n \in C^n$ with the property described above for each $q \in Q_A$ and $s_1 \ldots s_n\in A^n$, and define  $tr(q, s_1\ldots s_n) := s_1 c_1 \ldots s_n c_n$.
We set
\begin{equation*}
 R(q) := \{ tr(q, w) \mid w \in A^n \}.
\end{equation*}
Note that every $R(q)$ is a response, and for every $w \in R(q)$, $q$ ends up in $\top$ following a prefix of $w$. We say that a response with this property
\emph{leads} $q$ to $\top$.

We construct now a sequence $R^{(0)},R^{(1)},\ldots, R^{(m)} $ using the following algorithm.
Set $R^{(0)} := \{\epsilon\}$ and $i := 1$.
\begin{enumerate}
\item Pick a $q'_i \in Q_A$ that does end up in a node $q_i \not= \top$ following a $w \in R^{(i-1)}$. If no such $q_i$ exists set $R := R^{(i-1)}$ and terminate.
\item Set $R^{(i)} := (R^{(i-1)} \setminus \{w\}) \cup {w \circ R(q_i)}$. Set $i := i+1$ and go to 1).
\end{enumerate}

We show that for every $i$, if $w \in R^{(i)}$, $\abs{w} \leq n^2$. This implies termination of the algorithm.

Let $w \in R^{(i)}$.
Let $q'_1, \ldots, q'_m$ be the nodes selected in part (1) of the algorithm such that
$w = w_1 w_2 \ldots w_m$ with $w_j \in R(q'_j)$ for $1 \leq j \leq m$.
We define a family of node sets by:
\begin{itemize}
 \item $Q(0) = Q_A \cup \{\top\}$,
 \item for every $j \geq 1$, $Q(j)$ is the set of nodes consisting of $\top$ and all nodes $\hat{q}$ such that a $q \in Q_A$ ends up in $\hat{q}$
  following $w_1 \ldots w_j$.
\end{itemize}

For every $j \geq 0$, $\abs{Q(j)} \geq 1$.
We now prove that $Q(j)$ contains at most $n-j$ nodes. This is true for $Q(0)$.
For $j > 0$, note that $q'_j$ is chosen such that $w_j$ or one of its prefixes leads a $q \in Q(j-1)$ to $\top$. That implies
$\abs{Q(j)} < \abs{Q(j-1)}$, and therefore the property (recall that every node ends up in at most one node following a sequence).

Thus $m$ has to be smaller than $n-1$, and $\abs{w} \leq n^2$, since $R(q)$ has length $n$ for all $q$.
Note that for every $w, w' \in R^{(i)}$ for all $i \geq 0$, if $w \not= w'$ then $w|_A \not= w'|_A$.
Hence after termination of the procedure, we can replace every
$w \in R$ such that $\abs{w} = k\cdot n < n^2$ by $w \circ R'$, with $R'$ an arbitrary
response of length $(n-k) \cdot n$, to obtain equal length of all words in $R$, which then forms a response of length $n^2$.

For every $w \in R$, every node of $Q_A$ ends up in $\top$ after following a prefix of $w$.
We can conclude that every run $r$ with $\bar{r}|_G$ a prefix of a word in $(AC)^* R (AC)^\omega$ is terminating, and thus $(AC)^* R (AC)^\omega$ is a terminating pattern.
\qed
\end{proof}

}

\end{document}